\documentclass[12pt,preprint]{aastex}

\newcommand{\kms}{km~s$^{-1}$}
\newcommand{\lsim}{\hbox{ \rlap{\raise 0.425ex\hbox{$<$}}\lower 0.65ex\hbox{$\sim$} }}
\newcommand{\gsim}{\hbox{ \rlap{\raise 0.425ex\hbox{$>$}}\lower 0.65ex\hbox{$\sim$} }}

\slugcomment{Accepted to {\em AJ}}

\shorttitle{ORELSE I}
\shortauthors{Lubin et al.}

\begin{document}
\title{The Observations of Redshift Evolution in Large Scale
Environments (ORELSE) Survey : I. The Survey Design and First Results
on Cl 0023+0423 at $z = 0.84$ and RX J1821.6+6827 at $z = 0.82$}

\author{L.~M. Lubin\altaffilmark{1}, R.~R. Gal\altaffilmark{2},
B.~C. Lemaux, D.~D. Kocevski \& G.~K. Squires\altaffilmark{3}}

\affil{Department of Physics, University of California -- Davis, One
Shields Avenue, Davis, CA 95616} \altaffiltext{1}{lmlubin@ucdavis.edu}
\altaffiltext{2}{University of Hawai'i, Institute for Astronomy, 2680
Woodlawn Dr., Honolulu, HI 96822} \altaffiltext{3}{California
Institute of Technology, M/S 220-6, 1200 E. California Blvd.,
Pasadena, CA 91125}

\begin{abstract}

We present the Observations of Redshift Evolution in Large Scale
Environments (ORELSE) survey, a systematic search for structure on
scales greater than 10 $h_{70}^{-1}$ Mpc around 20 well-known clusters
at redshifts of $0.6 < z < 1.3$. The goal of the survey is to examine
a statistical sample of dynamically active clusters and large scale
structures in order to quantify galaxy properties over the full range
of local and global environments. We describe the survey design, the
cluster sample, and our extensive observational data covering at least
25$'$ around each target cluster. We use adaptively-smoothed red
galaxy density maps from our wide-field optical imaging to identify
candidate groups/clusters and intermediate-density large scale
filaments/walls in each cluster field. Because photometric techniques
(such as photometric redshifts, statistical overdensities, and
richness estimates) can be highly uncertain, the crucial component of
this survey is the unprecedented amount of spectroscopic coverage. We
are using the wide-field, multi-object spectroscopic capabilities of
the DEep Multi-Object Imaging Spectrograph to obtain 100-200+
confirmed cluster members in each field. Our survey has already
discovered the Cl 1604 supercluster at $z \approx 0.9$, a structure
which contains at least eight groups and clusters and spans 13 Mpc
$\times$ 100 Mpc. Here, we present the results on the large scale
environments of two additional clusters, Cl 0023+0423 at $z = 0.84$
and RX J1821.6+6827 at $z = 0.82$, which highlight the diversity of
global properties at these redshifts. The optically--selected Cl
0023+0423 is a four-way group-group merger with constituent groups
having measured velocity dispersions between 206--479 km s$^{-1}$. The
galaxy population is dominated by blue, star-forming galaxies, with
80\% of the confirmed members showing [OII] emission. The strength of
the H$\delta$ line in a composite spectrum of 138 members indicates a
substantial contribution from recent starbursts to the overall galaxy
population.  In contrast, the X-ray--selected RX J1821.6+6827 is a
largely-isolated, massive cluster with a measured velocity dispersion
of $926 \pm 77$ km s$^{-1}$. The cluster exhibits a well defined red
sequence with a large quiescent galaxy population. The results from
these two targets, along with preliminary findings on other ORELSE
clusters, suggest that optical selection may be more effective than
X-ray surveys at detecting less-evolved, dynamically-active systems at
these redshifts.

\end{abstract}

\keywords{catalogues -- surveys -- galaxies: clusters: general --
large-scale structure of the Universe}

\section{Introduction}

Clusters of galaxies are accurate tracers of the large scale structure
in the local Universe. Redshift surveys at $z < 0.1$ (e.g., Geller \&
Huchra 1989; Einasto et al.\ 1997; Colless et al.\ 2001) and numerical
simulations (e.g., Colberg et al.\ 2000; Evrard et al. 2002; Dolag et
al.\ 2006) reveal the filamentary structure of the Universe,
stretching between clusters and superclusters of galaxies.  Clusters
form at the nodes of these filaments, growing through the continuous
accretion of individual galaxies and groups from the surrounding field
(e.g., Frenk et al\ 1996; Eke et al.\ 1998).  The precursors of the
filaments should be present around distant clusters, containing many
of the galaxies which will eventually infall into the virialized core
and form the cluster population observed today. Since galaxy
environment should change dramatically during the course of vigorous
cluster assembly, the large scale structure present around
high-redshift clusters offers us the unique opportunity to probe, over
the full range of local environments, the physical effects on galaxies
as they assemble into denser regions.  As a result, we are undertaking
the Observations of Redshift Evolution in Large Scale Environments
(ORELSE) survey, a systematic search for structure on scales greater
than 10 $h_{70}^{-1}$ Mpc around 20 known clusters at $z > 0.6$.  The
survey covers 5 square degrees, all targeted at overdense cluster
regions, making it complementary and comparable to field surveys such
as DEEP2 (Davis et al.\ 2003) and COSMOS (Scoville et al.\ 2007;
Koekemoer et al.\ 2007).

While superclusters have been studied at low redshift (Davis et al.\
1980; Postman, Geller \& Huchra 1988; Quintana et al.\ 1995; Small et
al.\ 1998; Rines et al.\ 2002; Fadda et al.\ 2008; Porter et al.\
2008) and at $z \approx 0.2 - 0.6$ (Kaiser et al.\ 1999; Gray et al.\
2002; Kodama et al.\ 2001, 2003, 2005; Ebeling et al.\ 2004; Tanaka et
al.\ 2007a; Kartaltepe et al.\ 2008), large scale structures at high
redshift ($z \ge 0.6$) are just now being explored (Kodama et al.\
2005; Nakata et al.\ 2005; Tanaka et al.\ 2006, 2007b; Swinbank et
al.\ 2007; Gal et al.\ 2008; Fassbender et al.\ 2008; Gilbank et al.\
2008; Patel et al.\ 2008). Previous studies at these redshifts have
largely focused on the central regions ($< 2~h_{70}^{-1}$ Mpc) of
massive clusters. These studies have revealed strong evolution in the
cluster galaxy population which includes (1) an increase in the
fraction of blue, star-forming, late-type galaxies with redshift,
implying that early-types are forming out of the excess of late-types
over the last $\sim 7$ Gyr (e.g., Butcher \& Oemler 1984; Ellingson et
al.\ 2001; Dressler et al.\ 1997; van Dokkum et al.\ 2000, 2001; Lubin
et al.\ 2002); (2) a larger fraction of post-starburst (``K+A'')
galaxies in the cluster versus field environment, indicating strong
star formation activity in the recent past (e.g., Dressler et al.\
1999, 2004; Balogh et al.\ 1999; Tran et al.\ 2003); (3) a deficit of
faint, passive, red galaxies, suggesting that a large fraction of
these galaxies in present-day clusters has moved onto the red sequence
relatively recently as a result of a truncation of star formation
(e.g., Smail et al.\ 1998; van Dokkum \& Franx 2001; Kodama et al.\
2004; De Lucia et al.\ 2004, 2007; Tanaka et al.\ 2005, 2007b; Koyama
et al.\ 2007); and (4) an increase in the overdensities of X-ray and
radio sources with cluster redshift, indicating enhanced
starburst/nuclear activity in the past (e.g., Best 2003; Cappelluti et
al.\ 2005; Eastman et al.\ 2007; Kocevski et al.\ 2008a).

All of these findings imply significantly increased star-forming,
starburst, and nuclear emission in the past and raise the question --
what physical mechanisms associated with the cluster environment are
responsible for the suppression of star formation and nuclear activity
and the transformation of gaseous, disk galaxies into passive
spheroids?  Several mechanisms have been suggested, including galaxy
harassment (Moore et al.\ 1998), ram pressure stripping (Gunn \& Gott
1972), starvation (Larson et al.\ 1980), and merging (Mihos 1995,
1999). Most of these processes are associated {\it not} with the
densest cluster regions, but rather with the infall regions and
lower-density environments far from the cluster cores.

Studies at low-to-moderate redshift suggest that the non-cluster
processes play a pivotal role in driving galaxy evolution well before
the galaxies reach the cluster cores. Data from the Sloan Digital Sky
Survey (SDSS) and the 2dF Survey indicate a sharp transition between
galaxies with field-like star formation rates (SFRs) and galaxies with
low SFRs comparable to that of galaxies within the cluster cores
(Lewis et al.\ 2002; Gomez et al.\ 2003; Goto et al.\ 2003a,b; Balogh
et al.\ 2004). The density at which this transition occurs (log
$\Sigma \sim 1$ Mpc$^{-2}$) corresponds to the cluster virial radius,
well outside the core region on which most studies have been
focused. The fact that low SFRs and passive (gas-less) spiral galaxies
are observed well beyond the virial radius rule out severe processes,
such as ram pressure stripping or galaxy-galaxy merging, as being
completely responsible for the variations in galaxy properties with
environment (Lewis et al.\ 2002; Goto et al.\ 2003b). Similarly,
studies on large scales (clustocentric distances of $> 4~h_{70}^{-1}$
Mpc) at $z \sim 0.4$ indicate that galaxy colors change sharply at
relatively low densities and that the morphological mix, at a given
density, is independent of cluster radius, implying that galaxy
transformation occurs outside the cluster core and is physically
associated with the infalling filaments and the chains of smaller
galaxy groups within them (Kodama et al.\ 2001; Treu et al.\ 2003).
Consequently, these groups serve as a pre-processing phase in the
evolution of cluster galaxies (Kodama et al.\ 2001, 2003, 2004; Gray
et al.\ 2002; Treu et al.\ 2003; Bower \& Balogh 2004; Mihos 2004).

These trends continue to redshifts approaching $z \sim 1$ where there
are indications of ``downsizing'' with galaxy evolution accelerated in
high-density and high-mass (cluster) regions, compared to
lower-density and lower-mass (group) regions. Studies of large scale
structures at $z \sim 0.5-0.8$ suggest that the galaxy population in
clusters is more evolved, with the faint-end of the red sequence being
more developed and the red-sequence galaxies showing no signs of
current or recent star formation. In contrast, groups and
lower-density environments have a stronger deficit of the faint
red-sequence galaxies and clear signs of star-formation activity (weak
[OII] emission and/or strong H$\delta$ absorption) in the existing
red-galaxy population (De Lucia et al.\ 2004, 2007; Kodama et al.\
2004; Tanaka et al.\ 2005, 2005; Koyama et al.\ 2007). As such, these
results confirm that, at these epochs, much of the activity is taking
place in lower-density, lower-mass environments.

This environment--activity level connection also extends to the
extremes of active galactic nuclei (AGN) and starburst galaxies. From
studies covering a wide redshift range ($0.2 < z < 1.2$), excesses of
24 $\mu$m and X-ray sources are observed preferentially in
intermediate-density, not high-density, regimes, consisting of groups,
cluster infall regions, and filaments (Cappelluti et al.\ 2005;
Gilmour et al.\ 2007; Marcillac et al.\ 2007, 2008; Kocevski et
al. 2008b). These results, as well as those described above, imply
that the environmental processes which induce nuclear activity,
truncate star formation, and change galaxy properties are not
necessarily driven by cluster specific mechanisms, like ram pressure
stripping by the hot intracluster medium or harassment (i.e.,
truncation of the galaxy halo) by the cluster tidal field. Thus, it is
essential to look well beyond the regions of traditional study (the
cluster cores) to find answers to our questions concerning galaxy
evolution. The ORELSE survey is designed specifically to target these
largely-unexplored regions and answer these questions by correlating
multi-wavelength (radio, optical, infrared, and X-ray) photometric
data with galaxy kinematics and spectral features in a statistical
sample of large scale structures at redshifts approaching unity.

The first large scale structure detected in the ORELSE survey was the
Cl 1604 supercluster at $z \approx 0.9$ which includes two massive
clusters Cl 1604+4304 at $z = 0.90$ and Cl 1604+4321 at $z = 0.92$
originally detected by Gunn, Hoessel \& Oke (1986) and further studied
by Oke, Postman \& Lubin (1998). Wide-field imaging and spectroscopy
as part of the ORELSE survey has revealed a more complex, massive
structure, containing at least eight groups and clusters and spanning
13 Mpc $\times$ 100 Mpc (Lubin et al.\ 2000; Gal \& Lubin 2004; Gal,
Lubin \& Squires 2005; Gal et al.\ 2008). Our extensive spectroscopic
data on the Cl 1604 supercluster (over 400 confirmed members)
demonstrate that comprehensive redshift surveys, like ORELSE, are
essential for understanding galaxy and cluster evolution.
Specifically, (1) the entire structure size in redshift space is
equivalent to typical photometric redshift errors (Margoniner \&
Wittman 2008); (2) superpositions of groups/clusters mean that mass
measures based on weak lensing signal, richness, or X-ray luminosity
are highly uncertain; (3) cluster velocity dispersions based even on
traditionally large numbers of galaxies (i.e.\ 20-40) can be
substantially overestimated due to outliers (Gal et al.\ 2008); and
(4) the expected overdensities of radio and X-ray sources are small,
making the identification and study of individual active galaxies
impossible (Kocevski et al.\ 2008a).

In this paper, we build on our studies of the Cl 1604 supercluster. We
present the experimental design of the full survey (\S 2) and the
photometric and spectroscopic results on two additional target clusters,
Cl 0023+0423 (hereafter Cl 0023) at $z = 0.84$ and RX J1821.6+6827
(hereafter RX J1821) at $z = 0.82$, which highlight the significant
diversity of structure properties at these redshifts (\S 4 and
5). Throughout the paper we use a cosmology with $H_0=70$
$h_{70}^{-1}$ km s$^{-1}$ Mpc$^{-1}$, $\Omega_m=0.3$ and
$\Omega_{\Lambda}=0.7$.

\section{The Survey Design}

The ORELSE survey is a systematic search for structure on scales
greater than 10 $h_{70}^{-1}$ Mpc around 20 well-studied clusters at
redshifts of $0.6 < z < 1.3$ (see Table~\ref{sample}). The goal of the
ORELSE survey is to study the large scale structure around a
statistical sample of high-redshift clusters. Numerical simulations
indicate that 50\% of all $M > 3 \times 10^{14}~h^{-1}_{70}~M_{\odot}$
clusters at $z \sim 0.8$ have two or more companion clusters of equal
or greater mass within 100 co-moving Mpc (i.e., a configuration
similar to the Cl 1604 supercluster), and 70\% have had its most
recent large merger ($M_{final}/M_{initial} \ge 1.2$) within the last
$\sim 1$ Gyr (Colberg et al.\ 2000; Evrard et al.\ 2002; Cohn \& White
2005). Thus, our survey should identify at least ten structures like
the Cl 1604 supercluster and detect significant formation activity in
the majority of clusters in our sample.

\subsection{The Cluster Sample}

The cluster sample is chosen from various surveys (Oke, Postman \&
Lubin 1998; Rosati et al.\ 1998; Vikhlinin et al.\ 1998; Gladders \&
Yee 2000; Gioia et al.\ 1999, 2003; Stanford et al.\ 2002; Blanton et
al.\ 2003; Hashimoto et al.\ 2005; Henry et al.\ 2006; Pierre et al.\
2006; Maughan et al.\ 2006) with different detection techniques
(optical, X-ray, radio) to reduce any bias associated with sample
selection. Table~\ref{sample} lists the cluster coordinates, redshift,
detection technique, availability of high-quality (Chandra and/or XMM)
X-ray data, status of the optical/near-infrared observations, some
notes on the cluster properties, and original references. We note that
some of the target clusters, in particular the highest redshift ones,
are the targets of wide-field studies by other groups, including the
PISCES Survey (Kodama et al.\ 2005). We have included them in our
survey because the number of well-studied clusters at these redshifts
is still limited and because there exists a wealth of publically
available ground (Subaru and CFHT) and space-based (HST, Chandra, XMM,
and/or Spitzer) observations on these systems. Thus, we can increase
our sample size at very little observational cost to us.

The cluster sample also includes Cl 1604+4304 at $z = 0.90$ and Cl
1604+4321 at $z = 0.92$, the original clusters detected in the Cl 1604
supercluster (see Gal et al.\ 2008 and references therein). As the
first large scale structure detected in the ORELSE survey, the Cl 1604
supercluster currently has the most extensive multi-wavelength
photometric and multi-object spectroscopic data.  As discussed below,
we use the properties of the 419 confirmed members in the Cl 1604
supercluster as a guide when choosing the appropriate galaxy magnitude
and color selections in our other cluster fields.

\subsection{The Observations}

To detect large scale structure, we need to probe scales greater than
10 $h_{70}^{-1}$ Mpc, the mean projected separation of cluster pairs
in local superclusters (Bahcall et al.\ 1988).  Thus, the first phase
of our survey is wide field (25$'$) optical/near-infrared
($r'i'z'K_s$) imaging of the 20 cluster fields. All of the
near-infrared observations are original to the ORELSE survey, while
the optical imaging comes from both new and archived observations (see
Table~\ref{sample}). For our original observations, we are using the
Large Format Camera (LFC) and the Wide-Field Infrared Camera (WIRC) on
the Palomar 5-m, Suprime-Cam on the Subaru 8-m, the Wide-Field
Infrared Camera (WFCAM) on the UKIRT 3.8-m, and the Florida
Multi-object Imaging Near-IR Grism Observational Spectrometer
(FLAMINGOS) on the KPNO 4-m to cover an area of at least 0.2 deg$^2$
around each target cluster. The optical imaging is initially used to
characterize the large scale environment around the target clusters
and choose probable cluster members for follow-up multi-object
spectroscopy. The near-infrared $K_s$ imaging will be used to measure
stellar masses and provide improved photometric redshifts. The
photometric observations reach typical depths of $\{r'~i'~'z'~K_s\} =
\{24.7~24.2~23.4~20.5\}$ for a $3\sigma$ detection.

Based on the optical ($r'i'z'$) imaging for each cluster field, we use
broad color-color cuts, characteristic of red, elliptical-like
galaxies at the cluster redshift (see \S \ref{colorcut}), to produce
adaptively smoothed galaxy density maps. These maps are used primarily
to provide a visual locator for companion clusters and
intermediate-density large scale structures in the imaging field (see
\S \ref{dens}). Filaments and cluster infall regions can cover
significant portions of the observed area, but at relatively low
contrast, making it difficult to detect them and define their
boundaries.  Identification of such overdense regions is used to guide
the placement of follow-up observations, especially slitmasks for
multi-object spectroscopy.

Because photometric techniques for studying galaxy and cluster
evolution (such as photometric redshifts, statistical overdensities,
and richness estimates) are highly uncertain, the crucial component of
this survey, and what distinguishes it from competing studies, is the
unprecedented amount of spectroscopic coverage. We utilize the
wide-field, multi-object spectroscopic capabilities of the DEep
Multi-Object Imaging Spectrograph (DEIMOS; Faber et al.\ 2003) on the
Keck 10-m. Because of its large FOV ($16.7' \times 5.0'$), capability
of positioning up to 120+ galaxies per slitmask, and high efficiency,
DEIMOS is the ideal instrument for a spectroscopic survey of
high-redshift large scale structures. We perform faint object
spectroscopy (for objects down to $i' \approx 24$) with total exposure
times of 2.5--3 hrs. Using the density map as a guide (see above) and
a series of appropriate color cuts to preferentially select galaxies
at the cluster redshift (see \S \ref{speccut}), we are obtaining high
resolution ($\sim 60$ km s$^{-1}$) spectra for 100--200+ confirmed
members per system.  Our sample sizes are larger than most of those
currently available (e.g., Tanaka et al.\ 2006, 2007b; Koyama et al.\
2007; Swinbank et al.\ 2007; Fassbender et al.\ 2008) and allow us to
measure properties on a galaxy-to-galaxy basis. Based on the
spectroscopy completed so far (see \S3 and Gal et al.\ 2008), we find
that 25--45\% of the galaxies targeted through our system of color
selections are at or near the cluster redshift.

Based on our imaging and spectroscopic observations, promising
supercluster candidates and other complex systems are currently being
targeted for multi-wavelength studies with HST, Chandra, Spitzer,
and/or the VLA. We note that our large spectroscopic databases are
important to capitalize on the multi-wavelength data as we can begin
to identify and study rare galaxy populations (like X-ray and radio
sources) on an individual rather than a statistical basis (see
Kocevski et al.\ 2008a,b).

\subsection{The Color Selection}

To characterize the large scale structure around our target clusters,
we need to identify galaxies (which will be used to create a density
map and which will be targeted for spectroscopy; see below) within the
wide-field imaging that are at or near the cluster redshift. To do
this, we target red, elliptical-like galaxies which provide the best
contrast against the large background of blue field galaxies. Until we
have sufficient spectroscopy in each field to refine the color
selection, we need to choose initial color cuts in the $r'i'z'$ data
which are appropriate for the cluster redshift. We have examined the
possibility of applying color cuts based on the stellar population
synthesis models of \citet[][hereafter BC03]{bru03}. To see the
variation in colors between different star formation histories, we
generate synthetic galaxy spectra using three solar-metallicity models
: (1) an instantaneous burst with a formation redshift of $z_f = 3$;
(2) a $\tau=0.6$ Gyr exponential decline at $z_f = 3$; and (3) a
$\tau=0.6$ Gyr exponential decline at $z_f = 5$.  In Gal et al.\
(2008), we also explored the prescription used by the Red-sequence
Cluster Survey \citep[RCS,][]{gla05}, a model parameterized as a 0.1
Gyr burst ending at $z=2.5$, followed by a $\tau=0.1$ Gyr exponential
decline. We do not examine the RCS model here as it is nearly
identical (to within 0.04 mag) to the simpler model (\#1) for the
redshift range in which we are interested ($0.6 < z < 1.3$).

\begin{figure}
\plotone{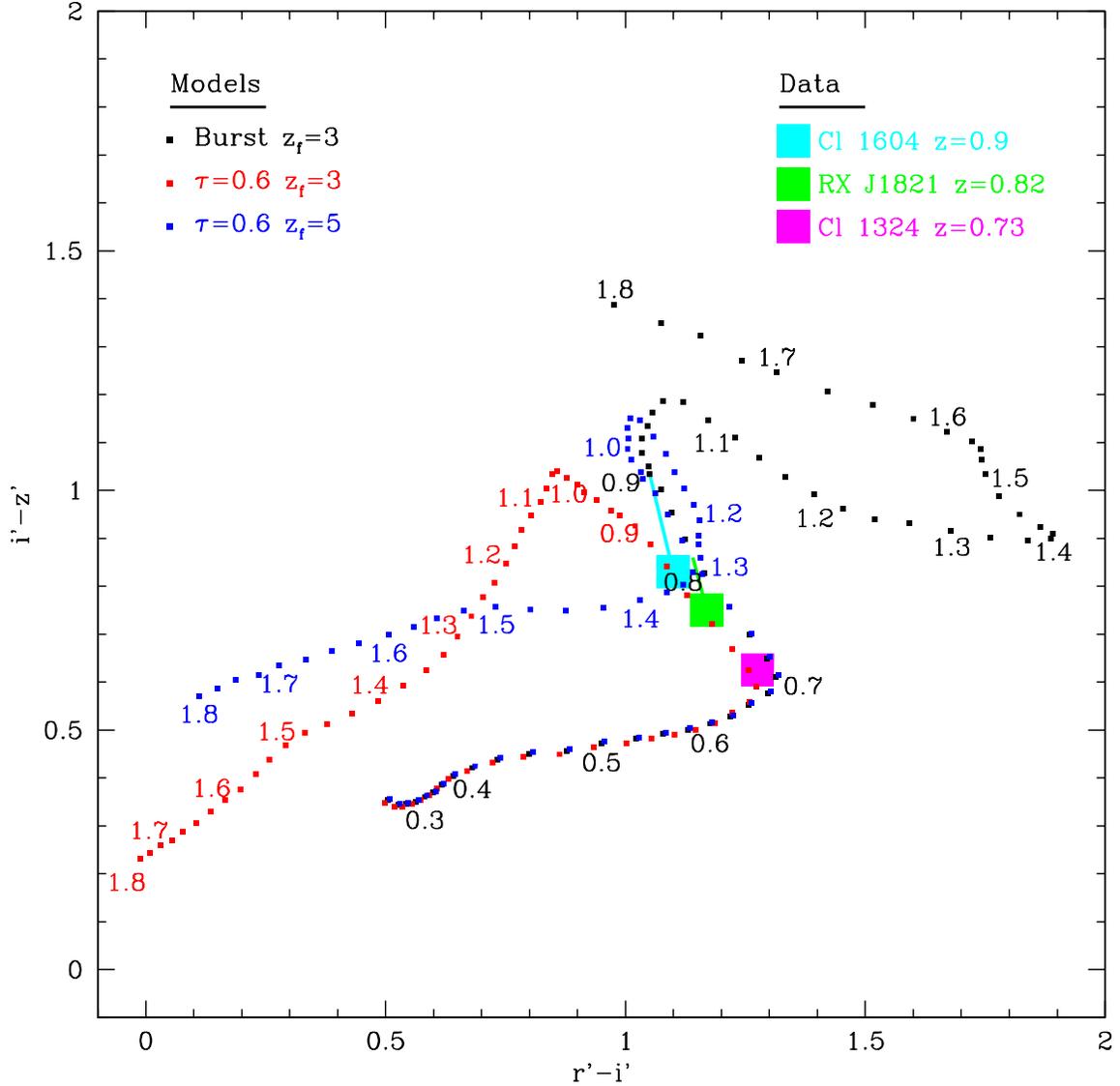}
\caption{The $r'-i'$ vs.\ $i'-z'$ color-color plane. The evolution
with redshift for three solar-metallicity BC03 models are shown (see
\S \ref{colorcut}) as dotted lines with the colors at a given redshift
indicated. Large filled squares represent the average colors of
spectroscopically-confirmed red-sequence galaxies in three ORELSE
structures, the Cl 1604 supercluster at $z \approx 0.9$ (cyan), the Cl
1324 supercluster at $z \approx 0.73$ (magenta), and RX J1821 at $z =
0.82$ (green). Solid lines show the offset from the observed colors of
these three systems to the colors predicted by the instantaneous burst
model at $z_f = 3$. No single model fits all three systems.}
\label{ccd} 
\end{figure}

All three models reproduce the colors of local elliptical galaxies
(e.g., Blanton et al.\ 2003; Chang et al.\ 2006) and have formation
epochs implied by the observed color evolution in red-sequence
galaxies (e.g., Stanford et al.\ 1998; van Dokkum \& Franx 2001;
Blakeslee et al.\ 2006; Mei et al.\ 2006b; Homeier et al.\ 2006). The
evolution with redshift in the ($r'-i'$) vs.\ ($i'-z'$) color-color
diagram (CCD) is shown in Figure~\ref{ccd} for each of the three
models. The first thing to note is that all three models are identical
at $z \le 0.6$ and nearly identical at redshifts of $0.6 < z < 1$,
suggesting that the color selection at these redshifts is insensitive
to the model as long as the formation epoch is sufficiently high. We
do, however, observe significant variations at $z > 1$, clearly due to
differences in formation redshift and starburst length. We note that
the abrupt changes in color with redshift, most notably in the
instantaneous burst model (\#1), are due to strong FeII and MgII
absorption lines shifting in and out of the optical bandpasses.

Overlaid in Figure~\ref{ccd} are the average colors of
spectroscopically confirmed red-sequence galaxies in three ORELSE
structures, the Cl 1604 supercluster at $z \approx 0.9$ (Gal et al.\
2008), the Cl 1324 supercluster at $z \approx 0.73$ (Gal et al.\ in
prep), and RX J1821.6+6827 at $z = 0.82$ (see Figure~\ref{52cmd}). We
do not include Cl 0023 here as it is dominated by blue galaxies and
has a wider red sequence than the other systems (see \S
\ref{galaxy}). This figure demonstrates that a single model does not
fit all three systems. In particular, the colors from the Cl 1604
supercluster and RX J1821.6+6827 are redder in $r'-i'$ and bluer in
$i'-z'$ than all of the models. Such discrepancies are unsurprising as
the evolutionary history of these systems are undoubtedly different
and more complex. In addition, a galaxy's star formation history will
vary with environment. As a result, we adopt broad ($\Delta m = 0.4$)
color cuts that were used successfully in the Cl 1604 supercluster. We
use the models as a guide and offset our color cuts relative to those
used in Cl 1604 (see \S 2.2 of Gal et al.\ 2008). For cluster
redshifts between $z = 0.6-1.0$, we use a standard $r'-i'$ cut of
1--1.4 but vary the $i'-z'$ cut depending on the cluster redshift.  As
done for the Cl 1604 supercluster, we adopt an $i'-z'$ cut of 0.6--1.0
for a cluster at $z = 0.9-1.0$. For those clusters at lower redshifts,
we adopt an $i'-z'$ cut which gets progressively bluer by 0.1 mag as
the redshift interval decreases by $\Delta z = 0.1$.
For cluster redshifts of $z > 1$, we use a fixed cut of $i'-z' >
0.9$. We use a single color cut at these redshifts because, while the
$r'-i'$ color at $z > 1$ is highly dependent on the model, the $i'-z'$
colors are consistently above $\sim 0.9$, except for models with
recent or continuing star formation. We note that ACS observations of
three X-ray--selected clusters at $z > 1$ all reveal well-defined red
sequences with characteristic colors of $i_{\rm F775W} - z_{\rm
  F850LP} \approx 1$ (Mei et al.\ 2006a,b; Demarco et al.\
2007). These results are completely consistent with our adopted
$i'-z'$ color cut. Table~\ref{ctab} lists the exact color cuts used in
this survey. \label{colorcut}

We note that our red-galaxy selection plus density-mapping technique
is fundamentally different from cluster detection methods such as the
red-sequence (RCS) method and MaxBCG \citep{koe07}. We use a simple,
relatively wide ($\Delta m \ge 0.4$ mag) color cut to examine the
projected density of objects with colors broadly consistent with those
of red galaxies at the redshift of interest. RCS and MaxBCG both rely
on the presence of a tight red sequence (with an assumed intrinsic
scatter of 0.075 and 0.060, respectively), which, especially at higher
redshift, may impose {\em a priori} requirements on the types of
systems detected. In the ORELSE survey, we target fields with known
structures and search for lower significance density enhancements,
such as groups and filaments, whose constituent galaxies, while more
evolved (redder) on average than field galaxies, do not necessarily
form a tight red sequence (as is the case for Cl 0023; see \S
\ref{galaxy}). While we have made broad assumptions on the formation
history of the cluster galaxies when choosing our color cuts
(particularly at $z > 1$) in order to eliminate fore/background
galaxies, the redshifts of our targets are already known, so we do not
require a photometric redshift estimator such as the red-sequence
fitting (which relies on one particular evolutionary model) used by
RCS.
  
\subsubsection{Magnitude Limits for the Density Map}

In addition to the color cuts used to select red galaxies at the
cluster redshift, we also need to define the range of $i'$ magnitudes
used to produce the density map (see below). In the case of the Cl
1604 supercluster at $z = 0.9$, we chose $i' = 20.5-23.5$ based on our
existing spectroscopy. For consistency, we shift the $i'$ magnitude
limits to the redshift of the other target clusters by using
$k$-corrections derived from the instantaneous burst at $z_f = 3$
model (\#1) shown in Figure~\ref{ccd}. The magnitude limits used to
create the red-galaxy density maps for the two target clusters
presented in this paper are given in \S \ref{pobs}.
\label{mlim}

\subsubsection{Priorities for Spectroscopic Observations}

When determining which galaxies to target for multi-object
spectroscopy, we prioritize galaxies based on color. As a first
priority, we select objects which meet our color cut(s) for red
galaxies described above. Within this priority (and the other lower
priorities), we weight objects by their $i'$-band flux with preference
given to brighter objects. We include objects with $i'$ between the
minimum magnitude used to create the density map (see \S \ref{mlim})
and $i' = 24$, the limit where we can still obtain reasonably high
signal-to-noise (S/N $\sim 1-5$) spectra in our 2.5--3 hr
exposures. With increasingly lower priority, we select galaxies which
are progressively bluer in both $r'-i'$ and $i'-z'$ to target galaxies
within the green valley and into the blue cloud. Although our
contamination rates are higher for bluer galaxies, we observe a
significant population of blue, star-forming members in the Cl 1604
supercluster (see Figure 1 of Gal et al.\ 2008) and Cl 0023 (see
Figure~\ref{00cmd}), so it is essential to target these galaxies as
well. We note that we always maximize the number of slits per slitmask
(given our minimum slit length of $5''$) by including all other
galaxies, within our magnitude limits, at the lowest priority. The
placement of these slitmasks (with the prioritized galaxy selection
described above) on the sky is guided by the structure (i.e.,
overdense regions and filaments) identified in the red-galaxy density
map (see \S \ref{dens}). \label{speccut}

\subsection{Producing the Density Map}

Following \citet{gal04}, we produce a red-galaxy density map for each
cluster field by adaptively smoothing a subset of galaxies which
satisfy the galaxy color selection and magnitude limits determined in
\S \ref{colorcut}.  An adaptive kernel smoothing is applied using an
initial window of $0.75~h^{-1}_{70}$ Mpc and 10 arcsecond pixels. The
chosen kernel size prevents small groups from being blended into
single detections in the density map and enhances the contrast of
small groups against the background, making detection of such low-mass
systems easier. We stress that the density maps of color-color
selected galaxies are primarily a tool for the visual identification
of structures such as clusters, groups, walls, and
filaments. \label{dens}

\subsubsection{The Cluster Detection Threshold} \label{detthr}

Because the ORELSE survey aims to detect comparatively low mass
systems, our cluster/group selection algorithm is potentially
susceptible to a high false detection rate. It is, therefore, useful
to select overdensities on the basis of some objective criteria.  The
application of both $r'-i'$ and $i'-z'$ color cuts to select galaxies
for the density maps results in a non-trivial relationship between
projected galaxy density and the cluster mass. A complete treatment of
selection functions would require large cosmological simulations with
model galaxy colors in our photometric system and would still depend
on the assumed cosmology and recipes for modeling galaxy
formation. Therefore, we rely instead on two observational datasets:
the extensive photometry and spectroscopy in the field of the Cl 1604
supercluster at $z=0.9$ \citep{gal04,gal08} for the cluster component
and the Canada-France-Hawaii Telescope Legacy Survey (CFHTLS) Deep
Fields\footnote{Based on observations obtained with MegaPrime/MegaCam,
a joint project of CFHT and CEA/DAPNIA, at the Canada-France-Hawaii
Telescope (CFHT) which is operated by the National Research Council
(NRC) of Canada, the Institut National des Science de l'Univers of the
Centre National de la Recherche Scientifique (CNRS) of France, and the
University of Hawaii. This work is based in part on data products
produced at TERAPIX and the Canadian Astronomy Data Centre as part of
the Canada-France-Hawaii Telescope Legacy Survey, a collaborative
project of NRC and CNRS.} for the ``field'' component.

The Cl 1604 system consists of numerous groups and clusters, with
velocity dispersions as low as $\sim 300$ km s$^{-1}$, corresponding
to a mass of $\sim 10^{14}~h^{-1}_{70}~{\rm M_{\odot}}$. The
spectroscopic catalog of over 400 confirmed supercluster members
allows us to reliably identify such low mass systems and examine their
properties in the map of galaxy density. We use the structure
identified as Cluster C in \cite{gal08} to set our detection
thresholds because it is the lowest-mass group to be spectroscopically
confirmed in the Cl 1604 supercluster.  This poor cluster is evident
as a single peak in the density map and has a sufficient number of
spectroscopic members (18 within 1 $h_{70}^{-1}$ Mpc) to estimate its
velocity dispersion $\sigma_C = 313$ km s$^{-1}$. Following
\citet{gal04}, we use SExtractor to detect structures in the density
maps. The SExtractor parameters must be chosen with care, to recover
systems which we know to be real (in this case, Cluster C and all of
the more massive systems in Cl 1604), while minimizing the number of
spurious detections.

To measure the false detection rate, we require a galaxy distribution
that retains the general correlation properties of galaxies meeting
our color-color selection criteria. Optimally, we would measure the
correlation function of field galaxies from the same data used for
cluster selection. However, our imaging survey is entirely targeted at
known high-density regions of the universe. Thus, we use the CFHTLS to
derive the statistical properties of the red galaxy distribution. The
2008A data release includes four fields, of $\sim2$ square degrees
each, at widely separately locations on the sky. Having four distinct
fields allows us to estimate not only the typical false detection rate
but also the range due to cosmic variance. Matched photometric
catalogs for all four fields are provided, including $r'i'z'$
photometry (the same as our data) in the AB system. In brief,
following \citet{pos96} and \citet{gal04}, we use the CFHTLS data to
compute Raleigh-Levy (RL) parameters for galaxies meeting our color
and magnitude cuts which are then used to generate simulated galaxy
distributions. These distributions are used to produce density maps,
on which we run SExtractor. The SExtractor parameters DETECT\_THRESH
and MIN\_AREA are varied until a set of values is found that
successfully detects the confirmed groups/clusters in Cl 1604, while
yielding low contamination.

In each CFHTLS Deep Field, we select those objects classified as
galaxies, meeting the same quality, $i'$ magnitude, and $r'-i'$,
$i'-z'$ cuts applied to the Palomar LFC data for the Cl 1604
imaging. This gives the number of galaxies $N_{RL,f}$ that should be
used in the RL simulation for the $f$-th CFHTLS Deep Field. This
number varies for every ORELSE target as the magnitude and color
limits change. A square region of exactly $1^{\circ}$ on a side is
selected from the center of each CFHTLS Deep Field to simplify
computation of the RL parameters and avoid underexposed field edges.
We then generate a set of simulated RL distributions over a grid of
values for the two RL parameters $\theta_0$ and $d$. The best match to
the observed CFHTLS data yields values of $\theta_0=100''$, $d=0.6$
using the color and magnitude cuts from Cl 1604. These parameters are
comparable to those found by \citet{gal04} and \citet{lop06}.

The next step is to create 100 simulated galaxy distributions for each
Deep Field using these RL parameters, each simulation having
$N_{RL,f}$ galaxies.  This yields 400 total density maps. We then run
the same adaptive kernel density mapper applied to the real data on
these simulated distributions and detect overdensities using
SExtractor. We vary the DETECT\_THRESH and DETECT\_MINAREA parameters,
using values of the galaxy density at different radii $r_{test}$ from
the center of Cluster C in Cl 1604, setting DETECT\_MINAREA to $\pi
r_{test}^2$.  We find that an area of 47 pix$^2$, corresponding to a
radius of 0.3 $h_{70}^{-1}$ Mpc, allows us to detect all of the
spectroscopically confirmed cluster candidates in the real data, while
minimizing the false candidates detected in the RL simulations. The
corresponding galaxy density threshold is 7300 galaxies per square
degree. The median number of false detections expected in the area
imaged for Cl 1604 is $N_{false}=0.6$, compared to the ten candidates
detected (see Figures 2--3 of Gal et al.\ 2008). The range of false
detections based on the four CFHTLS Deep Fields is between 0.3 and 0.9
for the Cl 1604 area.

As demonstrated in \citet{lop06}, the RL simulations provide a good
estimate of the false detection rate. A much simpler approach is to
generate purely random galaxy distributions, but these do not encode
any of the galaxy correlations present in the real universe.
Nevertheless, they serve as a useful confirmation of the RL
results. We performed the same analysis described above, using random
positions instead of an RL distribution. As expected, we detect many
fewer false candidates, with $N_{false}$ = 0.4 for the Cl 1604 area.

\subsubsection{False Detection Rates} 

Having set the SExtractor detection parameters based on Cl 1604, we
run cluster detection on the density maps for other cluster fields
with the same parameters. However, because the color and magnitude
limits vary from field to field based on the cluster redshifts and
pre-existing spectroscopy, the RL parameters and especially the number
of galaxies used in the RL simulation for each field will vary,
resulting in different contamination rates. First, we followed the
same procedure described above to determine the optimal RL parameters
for each ORELSE targets. For all such targets, we find that the range of
well-matched parameters is quite broad, so we fix the RL parameters to
$\theta_0=100''$, $d=0.6$ as found for Cl 1604.  Therefore, the only
quantity that we change in our RL simulation from target to target is
the total number of galaxies.

To determine the number of galaxies $N_{RL}$ to use for each ORELSE
target, we simply apply the $i'$ magnitude limits and $r'-i'$ and
$i'-z'$ color cuts used for the target to each of the four CFHTLS Deep
Fields.

For every target field $t$, we proceed as follows:

\begin{enumerate} 

\item We generate 100 RL distributions over each square CFHTLS Deep
Field with $N_{RL,t}$ galaxies. This number changes between the Deep
Fields due to cosmic variance.

\item Density maps are created from each of the distributions, with
 the same 0.75 $h_{70}^{-1}$ Mpc smoothing used for the corresponding
 target.

\item SExtractor is run with DETECT\_MINAREA corresponding to a radius of
  0.3 $h_{70}^{-1}$ Mpc at the redshift $z_t$ of the target, and
  DETECT\_THRESH fixed at the galaxy density from Cluster C in Cl 1604.

\item The number of expected false detections $N_{false,t}$ is set to
the mean value from the 400 simulations, scaled to the area imaged for
target $t$.



\end{enumerate}

The median number of false detections (and their range from the four
CHFTLS Deep Fields) expected in the fields of the two target clusters
discussed in this paper, Cl 0023 at $z = 0.84$ and RX J1821.6 at $z =
0.82$, is 0.60 (0.3--1.0) and 0.55 (0.3--0.8), respectively.
Cosmic variance on the scale of the LFC field could still potentially
cause significant changes in the number of false detections. We,
therefore, repeated the above process, now using 100 randomly placed
LFC-sized subfields within each of the 400 RL density maps for each
target. We find that the median and rms number of false detections is
$0.5\pm0.8$ and $0.3\pm0.6$ for Cl 0023 and RX J1821, respectively.
However, the median richnesses of these false detections is typically
much lower than for the overdensities found in our target fields (see
\S \ref{results}).

\section{The Targets and Observations}

\subsection{Cl 0023+0423}

Cl 0023 at $z = 0.84$ was originally detected as an overdensity in the
distant cluster survey of Gunn, Hoessel \& Oke (1984) which covered
71.5 deg$^2$ using photographic plates taken with the Palomar 1.2-m
and 5-m and the KPNO 4-m telescopes. This system was later included in
the Oke, Postman \& Lubin (1998) survey using the Low Resolution
Imaging Spectrograph (LRIS; Oke et al. 1995) on the Keck 10-m.  This
study spectroscopically confirmed 24 members and revealed that this
system is actually comprised of two small groups, with velocity
dispersions of 158 and 415 km s$^{-1}$ and separated by 2922 km
s$^{-1}$ in radial velocity (Postman, Lubin \& Oke 1998). N-body
simulations implied that the two groups were likely bound, were
currently merging, and would form a massive cluster ($\sigma_{los}
\approx 730$ km s$^{-1}$) in the next $\sim 1$ Gyr \citep{lub08b}.
Based on the LRIS photometry and spectroscopy, Postman, Lubin \& Oke
(1998) found that the population is quite active, with 57\% of the
confirmed members showing [OII] emission with equivalent widths
greater than 15 \AA\ and the majority with broad-band spectral energy
distributions (``color ages'') which suggest star formation within the
past 3 Gyr. Twelve confirmed members were covered by a single WFPC2
(F702W) image centered on this system. 75\% are late-type galaxies, a
morphological fraction more consistent with galaxy groups or the field
than rich clusters at these redshifts \citep{lub08a}.

Follow-up multi-wavelength studies of this system also indicate
substantial star formation and AGN activity. SCUBA observations of
four high-redshift ($0.84 \le z \le 1.27$) clusters, including Cl
0023, reveal 10 securely-identified submm sources detected toward the
cluster fields, a combined source excess at 850 $\mu$m of $\sim 3-4$
times blank field surveys (Best 2002). Similarly, an H$_{\alpha}$
imaging survey of Cl 0023 revealed a high integrated star formation
rate (SFR) per cluster mass of $(58.4 \pm 5.2)~h^{-3}$ M$_{\odot}$
yr$^{-1}$ / $10^{14}$ M$_{\odot}$, ten times higher than that observed
in H$_{\alpha}$ cluster surveys at $z \approx 0.2$
\citep{fin04}. \citet{fin04} also do not find a population of
early-type galaxies in Cl 0023 with on-going star formation (i.e.,
detectable H$_{\alpha}$ emission), suggesting that the [OII] emission
that we observe in otherwise red, passive galaxies may, in fact, be
due to AGN activity (see \S \ref{galaxy}).

\subsection{RX J1821.6+6827}

RX J1821 at $z = 0.82$ was the highest-redshift cluster discovered in
the ROSAT North Ecliptic Pole (NEP) survey (Gioia et al.\ 2003; Henry
et al.\ 2006). XMM observations revealed slightly elongated X-ray
emission, with a bolometric luminosity of $1.17^{+0.13}_{-0.18} \times
10^{45}~h_{70}^{-2}$ erg s$^{-1}$ and a temperature of
$4.7^{+1.2}_{-0.7}$ keV (Gioia et al.\ 2004).  Spectroscopic redshifts
for twenty cluster galaxies using the Keck I and the
Canada-France-Hawai'i (CFH) telescopes imply a cluster velocity
dispersion of $775^{+182}_{-113}$ km s$^{-1}$, typical of a relatively
rich cluster (Gioia et al.\ 2004).

The cores of all 18 clusters in the NEP survey were imaged by the VLA
at 1.4 GHz (Branchesi et al.\ 2006). A total of 32 radio sources, with
peak brightnesses $\ge 0.17$ mJy/beam, were detected, suggesting a
3$\sigma$ overdensity of sources relative to the field within 0.4
$h^{-1}_{70}$ Mpc. One of these sources was a confirmed member of RX
J1821. These results indicate radio AGN activity in the cluster cores,
consistent with studies of other clusters (Owen et al.\ 1999; Best
2003; Best et al.\ 2002; Miller \& Owen 2003).

\subsection{Photometric Data} 

Cl 0023 and RX J1821 were imaged using the Large Format Camera
\citep[LFC;][]{sim00} on the Palomar 5-m telescope. The LFC is a
mosaic of six $2048\times4096$ CCDs in a cross-shaped layout, mounted
at prime focus. The imaging area corresponds roughly to the
unvignetted region of a circle $24'$ in diameter. Data were taken on
UT 27 April and 15 August 2004, with the camera installed in the
standard North-South orientation and all six CCDs in use. The imaging
data were taken in unbinned mode, with a pixel scale of $0\farcs182$
pix$^{-1}$. Average seeing was $1\farcs0$, and the nights were not
photometric. Individual exposures of 450 sec were taken, moving the
telescope by up to $45''$ in each direction to alleviate the gaps due
to the spaces between the CCDs. Each target was imaged in the Sloan
Digital Sky Survey (SDSS) $r'i'z'$ filters, with total exposure times
in these filters of $\{4500~3600~2250\}$ sec for Cl 0023 and
$\{6300~4050~3600\}$ sec for RX J1821.

Data reduction was performed using the IRAF data reduction suite
\citep{tod86}, including the external package \textsc{mscred}.  We
followed the general mosaic imaging guidelines used by the NOAO Deep
Wide Field Survey \citep{jan04}, with modifications appropriate to our
data. The details of the data reduction are described in
\citet{gal05}. Object detection in the target frames was performed by
running SExtractor in dual-image mode, using ultra-deep images
(combining all three bands) for detection, while measurements were
performed on the single-band images. Elliptical photometry apertures
are determined from the deep image, and these same apertures are used
for the individual filter images. This procedure results in catalogs
where magnitudes are measured using identical apertures in all three
filters, which improves the measurement of galaxy colors by using the
same physical size for each galaxy in all filters \citep{lub00}. Due
to the lack of calibration, we assumed photometric zero points, color
terms, and atmospheric extinction terms from other photometric nights,
resulting in semi-calibrated catalogs.

To provide photometric calibration, RX J1821 was observed with the
Orthogonal Parallel Transfer Imaging Camera \citep[OPTIC;][]{how03}
installed on the University of Hawai'i 2.2-m telescope on UT 22 July
2007. This instrument contains two 2K $\times$ 4K CCDs mounted
side-by-side, with $0\farcs14$ pixels. While this instrument is
capable of orthogonal transfer charge shifting to improve image
quality, we used it in conventional mode to minimize data reduction
complexity. Three exposures of 300 sec each 
were obtained in each of the SDSS $r'$, $i'$, and $z'$ filters in
photometric conditions with $0\farcs8$ seeing. Numerous SDSS standard
stars were also observed to derive photometric calibration. These
observations were processed and sources detected the same way as for
our LFC data. The calibrated, photometric source catalogs from OPTIC
were matched to the semi-calibrated catalogs from LFC, and
transformations of the form
 \begin{equation}
m_{calib} = m_{LFC} + A\times(color) + B
\end{equation}
were derived for each filter, using the $r'-i'$ color for the $r'$ and
$i'$ filters, and $i'-z'$ color for $z'$. These transformations were
then applied to the full source catalogs from the LFC data.
Unfortunately, we have not yet been able to obtain photometric
calibration for the Cl 0023 field. Comparison of the semi-calibrated
LFC data of RX J1821 with the final calibration using OPTIC suggests
that zero points in each filter may be incorrect by 0.05--0.1 mag,
with errors of $\sim0.1$ on the color term.

The resulting color-magnitude diagrams (CMDs) and color-color diagrams
(CCDs) for the Cl 0023 and RX J1821 fields are shown in
Figures~\ref{00cmd} and \ref{52cmd}. Based on the photometry described
above, red-galaxy density maps were made using the appropriate
color-color cuts (see \S \ref{colorcut} and Table~\ref{ctab}) and
$i'$-band magnitude ranges of 20.28--23.28 and 20.16--23.16 for Cl
0023 and RX J1821, respectively (see \S
\ref{mlim}). Figures~\ref{00dens} and \ref{52dens} show the resulting
density maps. Using the RL simulation of \S \ref{dens}, we expect 0.60
and 0.55 false detections in the Cl 0023 and RX J1821 fields,
respectively, significantly fewer the actual number that we detect
(see \S \ref{results}).

\label{pobs}

\subsection{Spectroscopic Data} 

Both Cl 0023 and RX J1821 were observed with the Deep Imaging
Multi-object Spectrograph \citep[DEIMOS;][]{fab03} on the Keck 10-m
telescopes.  We used the 1200 l/mm grating, blazed at 7500 \AA, with
the OG550 filter and $1''$ slits. This setup results in a pixel scale
of 0.33 \AA~pix$^{-1}$, a resolution of $\sim$1.7 \AA\ (68 km
s$^{-1}$), and typical spectral coverage from 6250 \AA\ to 8900
\AA. Galaxies as faint as $i' \approx 24$ were observed, with highest
priority given to red galaxies based on the color cuts used to produce
the initial density maps from the semi-calibrated photometry.
Approximately 60\% of the targeted objects met the red galaxy color
cut. Lower priorities were given to galaxies increasingly blueward (in
$r'-i'$ and/or $i'-z'$) of the highest priority box. Cl 0023 was
observed with three masks and RX J1821 with two masks on UT 02-03
September 2005 in $0\farcs7$ seeing, with 5 exposures of 1800s each on
each mask. Cl 0023 was observed with two additional masks on UT 12
September 2007 with 5 exposures of 1800s each in $0\farcs8$ seeing and
thin cirrus.

Data were reduced using the DEEP2 version of the {\em spec2d} and {\em
spec1d} data reduction pipelines \citep{dav03}. Redshifts were
obtained using DEEP2's redshift measurement pipeline {\em zspec}
(Cooper et al.\ in prep), with typical redshift errors of
25~{\kms}. Each redshift was assigned a quality flag between 1 and 4,
where 1 indicates a secure redshift could not be determined due to
poor signal, lack of features, or reduction artifacts; 2 is a redshift
obtained from either a single feature or two marginally detected
features; 3 is a redshift derived from at least one secure and one
marginal feature; and 4 is assigned to spectra with redshifts obtained
from several high signal-to-noise features. A quality of -1 is used
for sources securely identified as stars.  Qualities of 3 or 4
indicate secure extragalactic redshifts and only those are used in the
following analyses (see also Gal et al.\ 2008).

In the Cl 0023 field, we obtained 423 extragalactic redshifts with
$Q\ge3$, of which 54 were serendipitous detections. An additional 73
galaxy redshifts were incorporated from the spectroscopic survey of
Oke, Postman \& Lubin (1998). The combined catalog contains $Q\ge3$
redshifts for 169 structure members with $0.82<z<0.87$. In the RX
J1821 field, a total of 129 extragalactic redshifts with $Q\ge3$ were
obtained in the RX J1821 field, 26 of which were serendipitous
detections. Thirteen additional redshifts were taken from
\citet{gio04}.  The combined catalog contains $Q\ge3$ redshifts for 85
structure members with $0.80<z<0.84$.

\section{The Results}

\label{results}

\subsection{The Global Properties}

In Figures~\ref{00dens} and \ref{52dens}, we show the red-galaxy
density maps for the Cl 0023 and RX J1821 fields. The density peaks
which meet our detection threshold (see \S \ref{detthr}) are
circled. We find three peaks in the Cl 0023 field and one in the RX
J1821 field, all significant based on our estimated false detection
rate.  The observed structures in these maps are consistent with that
expected from the redshift distributions shown in Figures~\ref{00hist}
and \ref{52hist}. While Postman, Lubin \& Oke (1998) found only two
peaks in the redshift distribution, we see four clear peaks with our
more extensive spectroscopy. Combined with the density map, these
results suggest an even more complex system of groups. In contrast,
the redshift histogram of RX J1821 shows a single peak at $z \approx
0.82$ (albeit with some clear velocity substructure; see below) which
is consistent with the single, massive overdensity observed in the
density map.

In addition, the distribution of confirmed members in each system
follows exceedingly well both the peaks and lower-density regions in
the two density maps. In the Cl 0023 field, although many of the
confirmed members are spread out across the spectroscopic area, 54\%
coincide directly (within 1 $h_{70}^{-1}$ Mpc) with the three
significant peaks (labeled A, B, and C). This correlation is not due
to spectroscopic coverage alone as the five DEIMOS masks cover a large
fraction of the imaging area, with most of that area covered by at
most two masks (see left panel of Figure~\ref{00dens}). We also see
that the density map does an excellent job at picking up associated
lower-density structures, such as the less-significant peak directly
to the South of component B. Splitting the confirmed members by
redshift according to the four redshift peaks in Figure~\ref{00hist},
we find velocity segregation between the three components. Component A
is largely comprised of galaxies belonging to the redshift peak at $z
\approx 0.839$, while component C is largely comprised of galaxies
belonging to the redshift peak at $z \approx 0.845$. Component B,
however, is actually a superposition of two groups along the
line-of-sight, one which corresponds to the lowest redshift peak at $z
\approx 0.828$ and the other which corresponds to the redshift peak at
$z \approx 0.845$, same as component C.  The galaxies within the
highest redshift peak at $z \approx 0.864$ are not centrally
concentrated but rather extend across the entire region, suggesting a
sheet of galaxies in the near background.

In Figure~\ref{00vel}, we show the redshift histograms, within radii
of 0.5 and 1.0 $h^{-1}_{70}$ Mpc, centered on the three overdense
components. Following the procedure described in Gal et al.\ (2008),
we use these distributions to measure the biweight velocity dispersion
computed by ROSTAT \citep{bee90}. Components A and C represent single
systems with measured velocity dispersions between 428--497 km
s$^{-1}$, typical of massive groups or poor clusters (e.g., Zabludoff
\& Mulchaey 1998). Component B is comprised of two smaller galaxy
groups with measured velocity dispersion of 206--293 km s$^{-1}$ (see
Table~\ref{candinfo}). As such, Cl 0023 is a four-way group-group
merger. Based on virial mass estimates of the individual groups, the
final mass of the merged cluster will be $\sim 5 \times 10^{14}~
h^{-1}_{70}$ M$_{\odot}$, typical of a rich cluster (e.g., Girardi et
al.\ 1998).

While Cl 0023 is dynamically active, RX J1821 is considerably more
relaxed with a single peak in both the density map and redshift
histogram (see Figure~\ref{52dens} and \ref{52hist}); however, there
are indications of continuing formation in this cluster as well. In
particular, the overdense region in the density map corresponding to
the cluster is clearly extended in the North-South direction,
consistent with the X-ray contours from Gioia et al.\ (2004). As in Cl
0023, the distribution of confirmed members follows extremely well the
overdense regions in the density map, tracing the elongated structure
of the cluster and even picking up the lower-density peaks (small
groups) to the South and South-west which are kinematically associated
with the cluster (right panel of Figure~\ref{52dens}).

To determine if RX J1821 has any velocity segregation in the plane of
the sky, 
we have binned the velocities by sky position going from North to
South along the elongated structure. We use four bins, each with an
equal number of galaxies.  Comparing each resulting velocity histogram
to the full velocity histogram using the KS test, only the
Southern-most distribution is marginally different, with a 6\%
probability of being drawn from the same distribution, possible due to
an infalling group (see Figure~\ref{52dens}). Otherwise, no
statistically significant velocity segregation was detected.

Velocity histograms, centered on the density peak and measured within
a projected radius of 0.5, 1.0, and 1.5 $h^{-1}_{70}$ Mpc, are shown
in Figure~\ref{52vel}.  We measure biweight velocity dispersions
between 926--993 km s$^{-1}$ (see Table~\ref{candinfo}). We note that
three ROSTAT dispersion estimators (sigma, biweight, and gapper) give
consistent measurements within a radial bin and from bin to bin;
however, as is clear from Figure~\ref{52vel}, the velocity
distributions are clearly non-gaussian, implying measurable velocity
substructure. RX J1821 does not, however, have clear multiple
components like Cl 0023 or a significant amount of nearby structure
like the Cl 1604 supercluster (Gal et al.\ 2008).

The variation in system properties might reflect the differences in
selection technique. Based on the handful of ORELSE targets observed
so far, we find that optical selection appears to preferentially
select systems which are dynamically active (Cl 0023), are embedded in
complex, large scale structures (the Cl 1324 and 1604 superclusters),
contain lower-mass constituents (all of the above), and have less (or
no) diffuse gas emission (all of the above; see Lubin, Mulchaey, \&
Postman 2004; Kocevski et al.\ 2008a; Kocevski et al.\ in prep). As
often suggested, X-ray selection appears to detect more
dynamically-evolved, concentrated, and isolated clusters (RX J1821 and
another NEP cluster, RX J1757.3+6631 at $z = 0.691$)
at these redshifts.

\label{global}

\subsection{The Galaxy Populations}

\begin{figure}
\plottwo{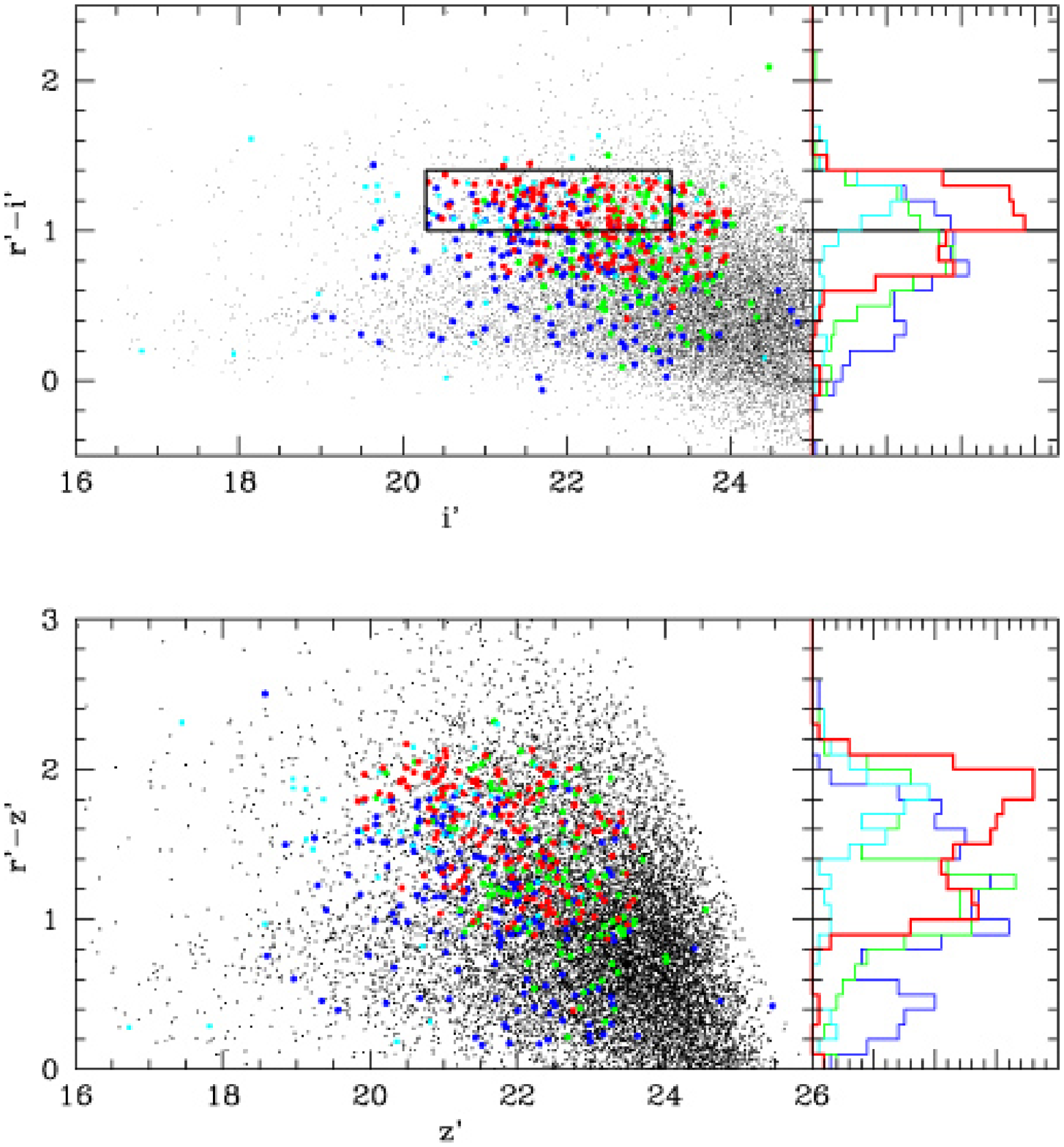}{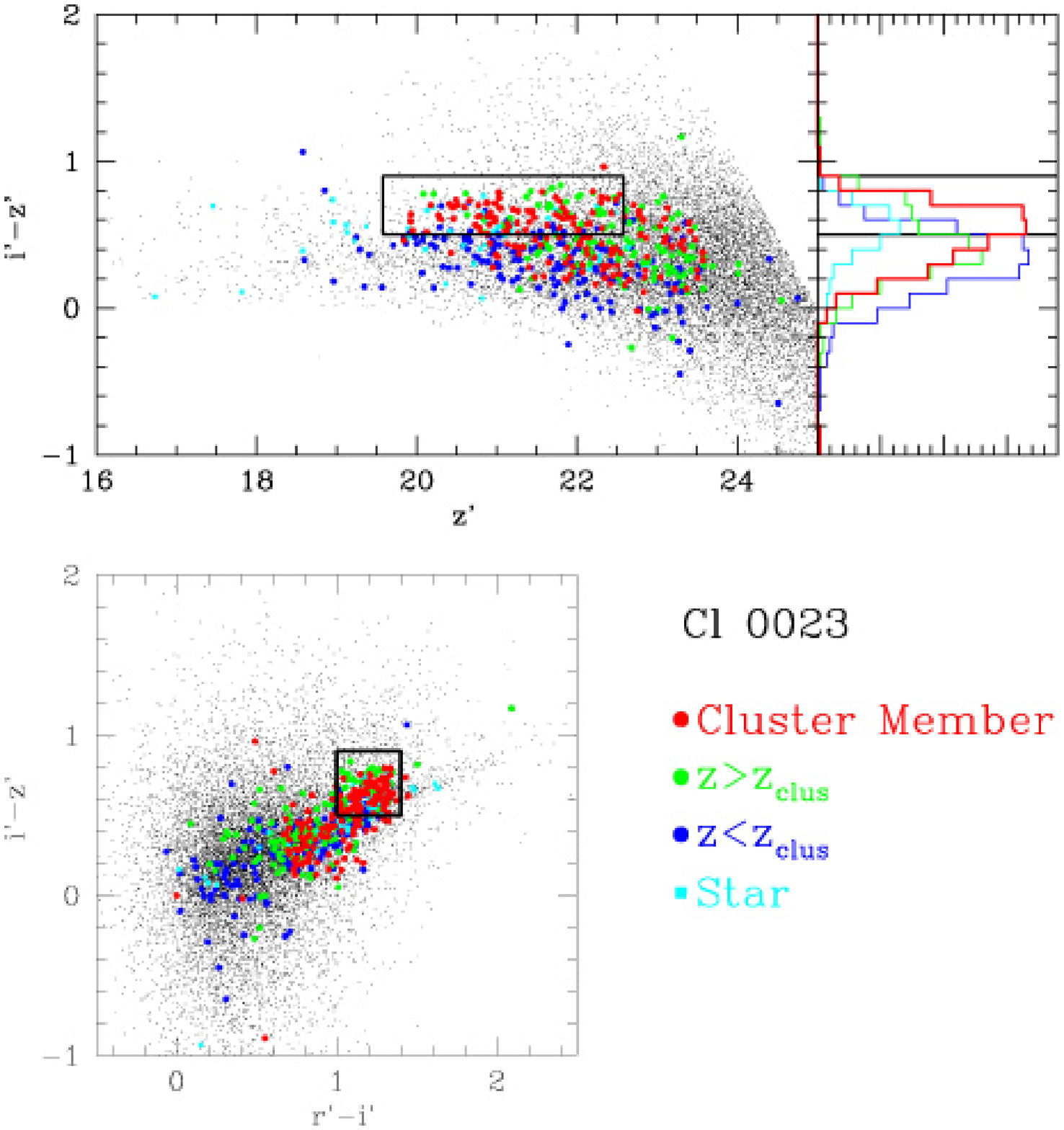}
\caption{The $i'$ vs.\ $(r'-i')$, $z'$ vs.\ $(r'-z')$, $z'$ vs.\
$(i'-z')$, and $(r'-i')$ vs.\ $(i'-z')$ color-magnitude and
color-color diagrams of the Cl 0023 field. All objects in the LFC
imaging are shown as small dots. The black rectangular regions outline
the color-selections applied to produce the density maps and
prioritize spectroscopic targets. Spectroscopically confirmed system
members are overplotted as large red dots. Foreground galaxies at
$z<0.82$ are shown as blue dots, background galaxies at $z>0.87$ are
green dots, and cyan dots are stars. Color distributions of
spectroscopic objects are shown to the right of each CMD.}
\label{00cmd} 
\end{figure}

\begin{figure}
\plottwo{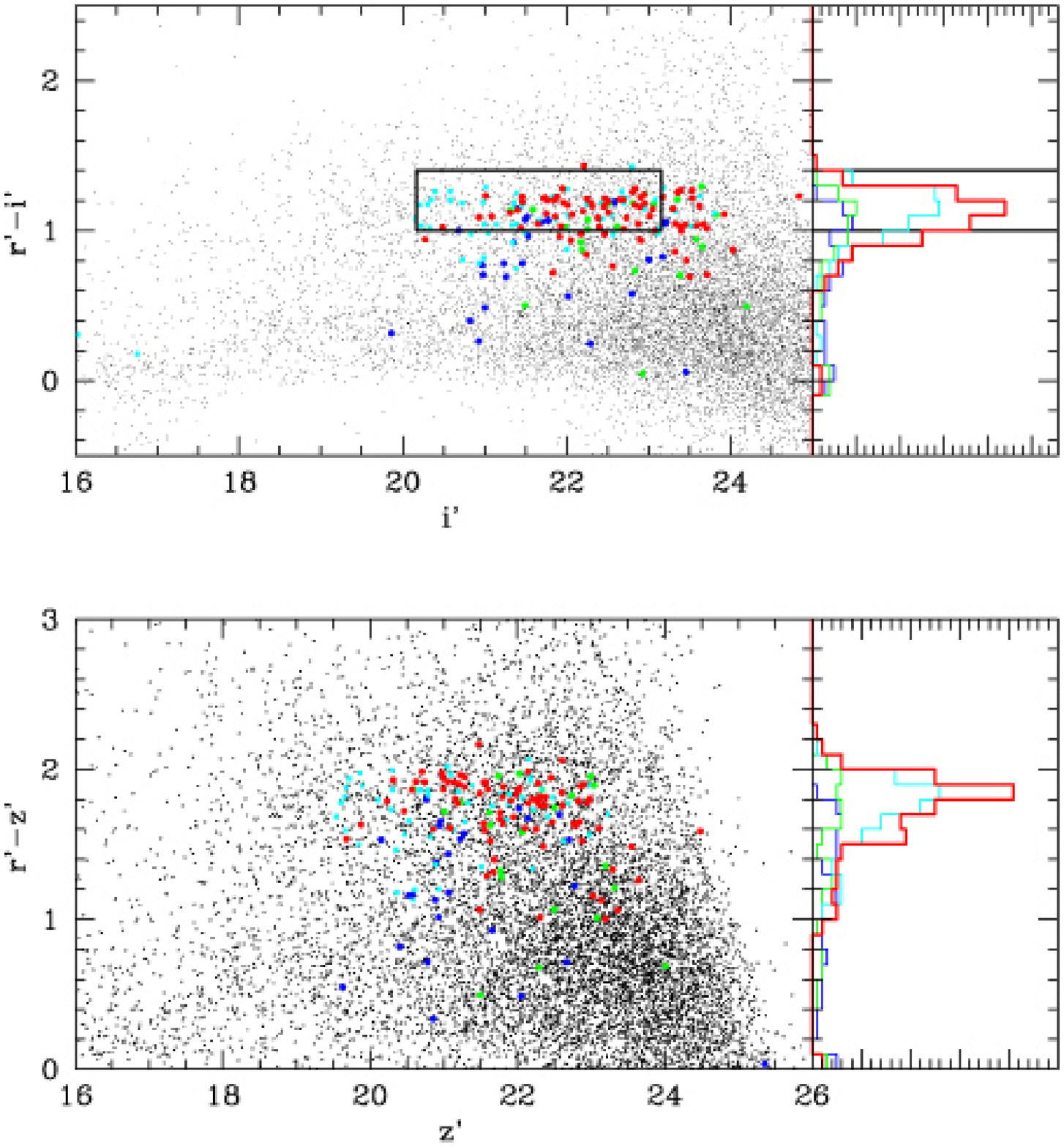}{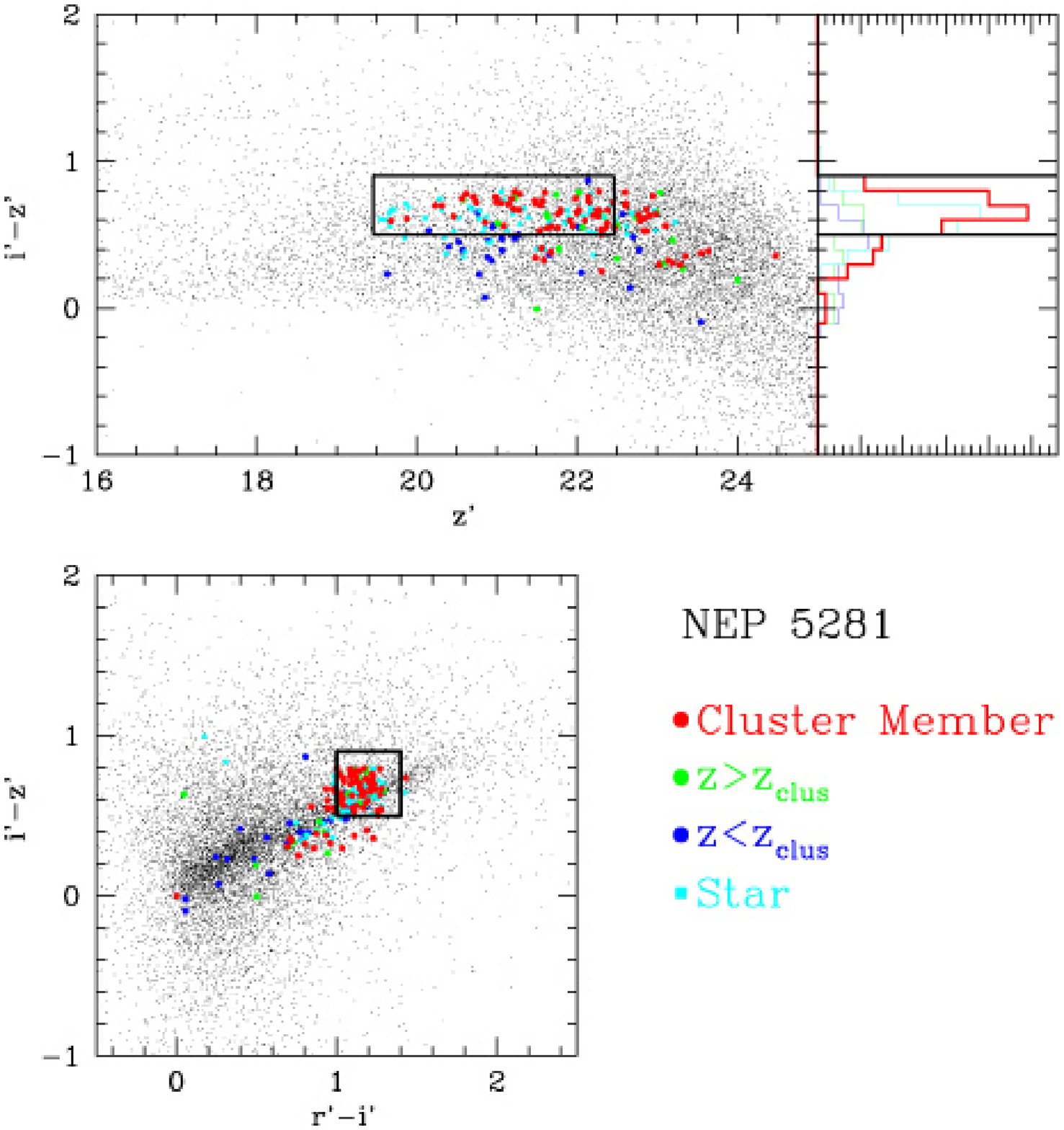}
\caption{Same as Figure~\ref{00cmd} but for the RX J1821
field. Spectroscopically confirmed cluster members are overplotted as
large red dots. Foreground galaxies at $z<0.80$ are shown as blue
dots, background galaxies at $z>0.84$ are green dots, and cyan dots
are stars.}
\label{52cmd} 
\end{figure}

\begin{figure}
\plottwo{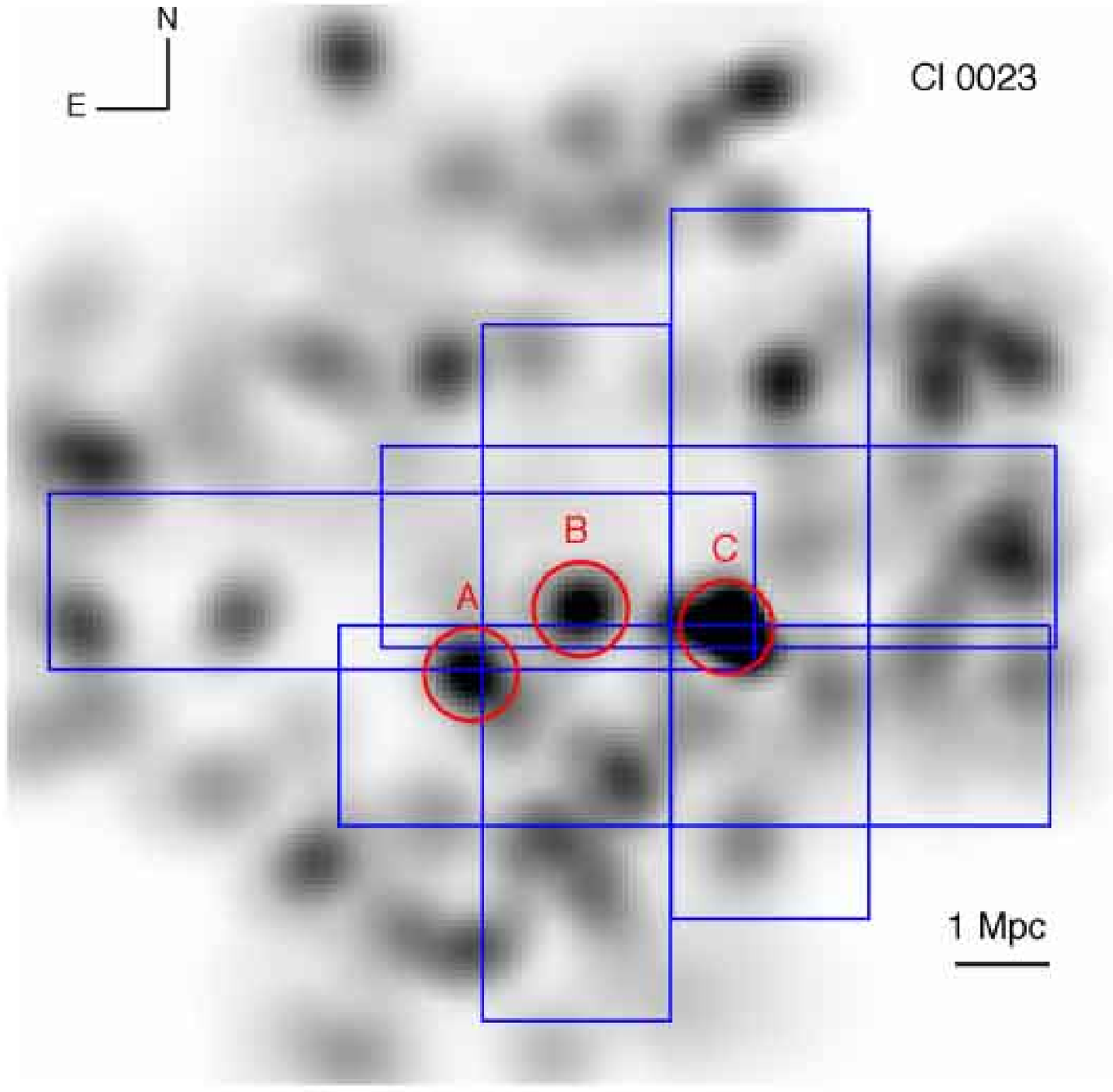}{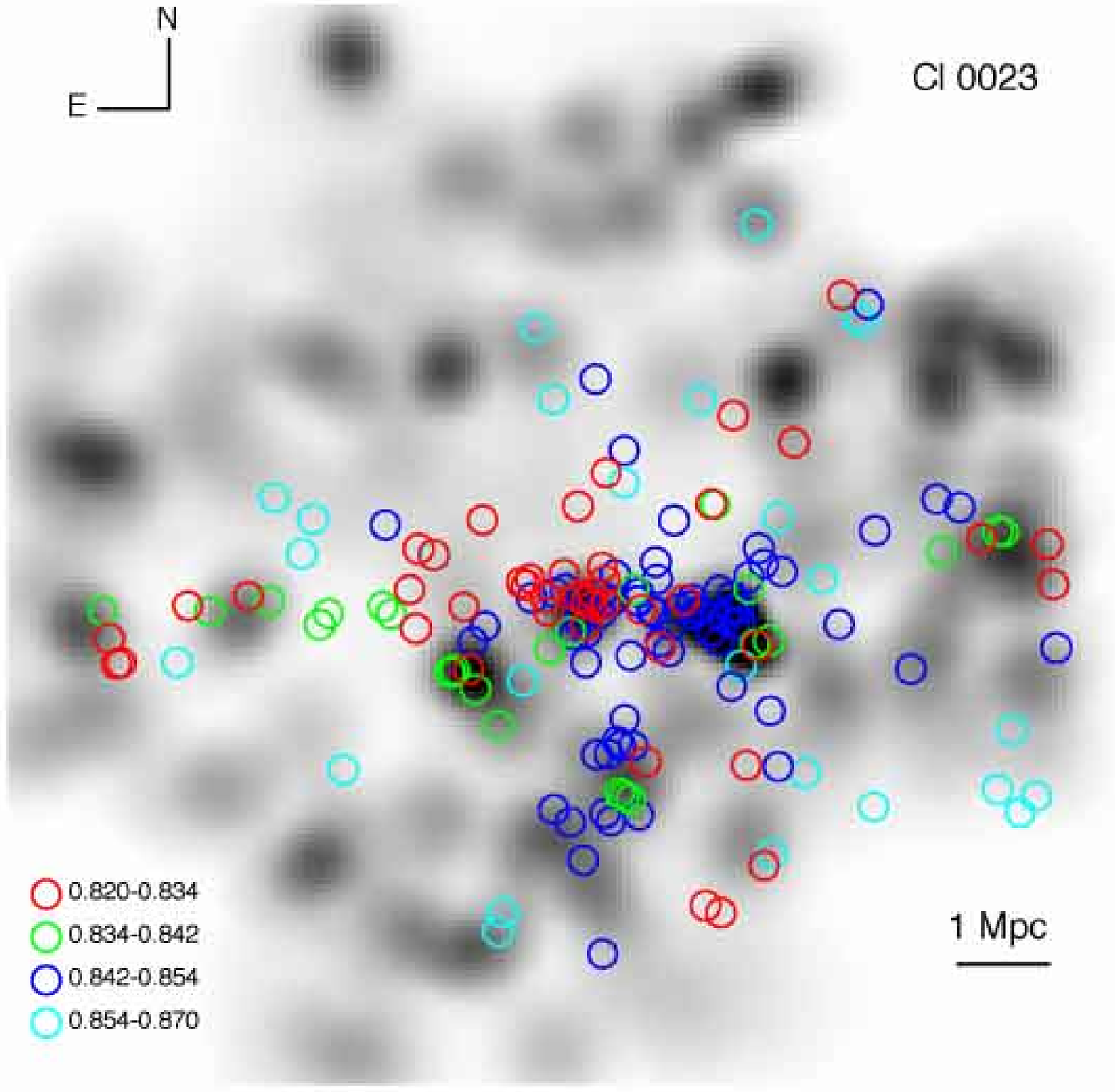}
\caption{Adaptive kernel density map of color-selected red galaxies in
the Cl 0023 field. ({\it Left}) Candidate groups/clusters, which meet
our detection threshold (see \S \ref{dens}), are marked with red
circles of $0.5~h^{-1}_{70}$ Mpc and labeled with an identifying
letter. Large blue rectangles indicate the layout of the five DEIMOS
masks. ({\it Right}) Small circles indicate the 169 confirmed members,
coded by color representing four redshift ranges which correspond to
peaks in the redshift distribution in Figure~\ref{00hist}. The
distribution of confirmed members corresponds extremely well to peaks
in the density map, including, as expected, the three significant
peaks but also the lower-density peak directly South of component B.}
\label{00dens} 
\end{figure}

\begin{figure}
\plottwo{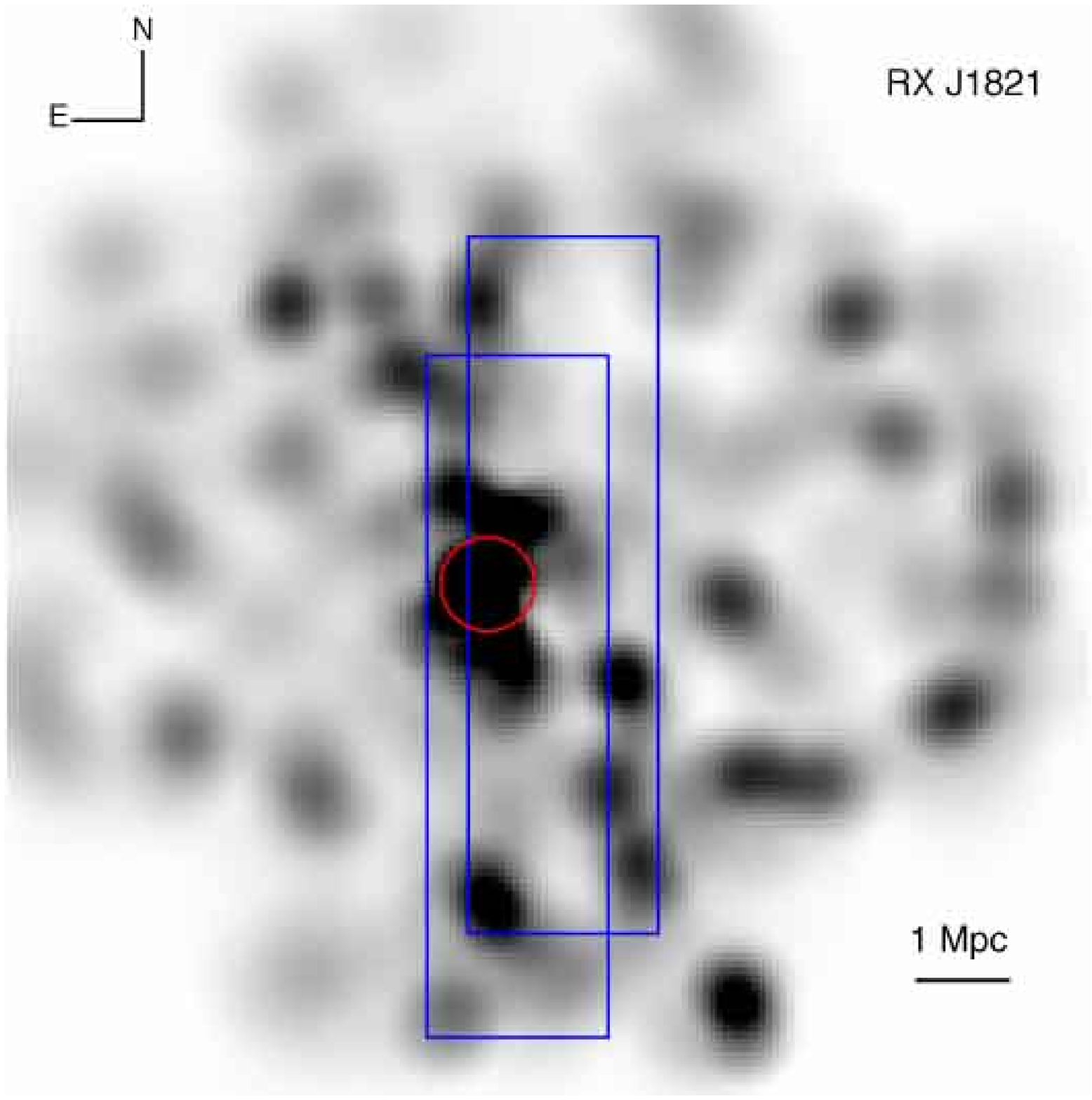}{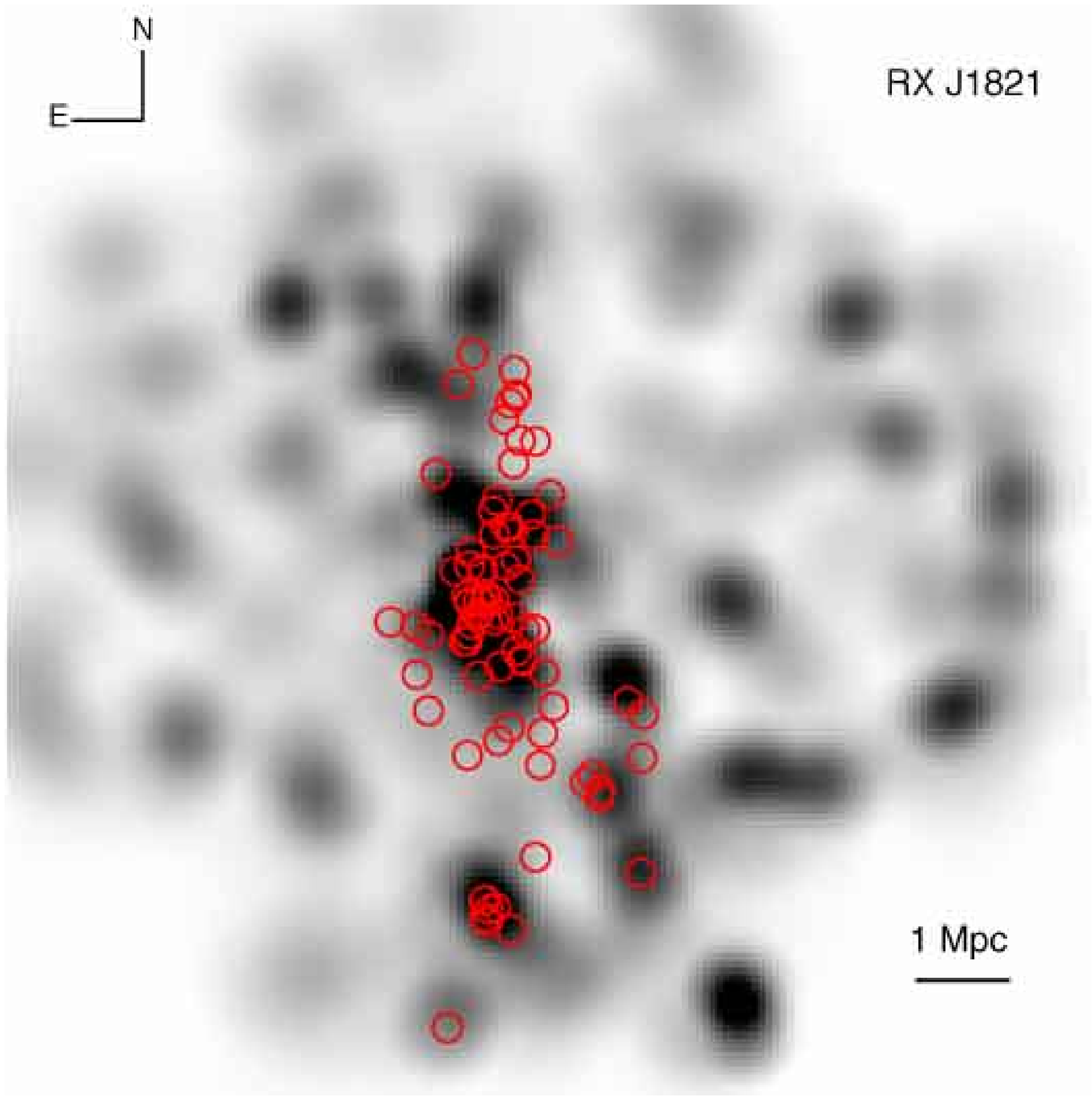}
\caption{Adaptive kernel density map of color-selected red galaxies in
the RX J1821 field. ({\it Left}) Only the cluster itself, marked by a
red circle of radius $0.5~h^{-1}_{70}$ Mpc, meets our detection
threshold.  Large blue rectangles indicate the layout of the two
DEIMOS masks. ({\it Right}) Small red circles indicate the 85
confirmed cluster members. As in Figure~\ref{00dens}, the distribution
of confirmed members follows extremely well the overdense regions in
the density map, tracing the elongated structure of the cluster and
even picking up the lower-density peaks (small groups) to the South
and South-west.}
\label{52dens} 
\end{figure}

\begin{figure}
\plotone{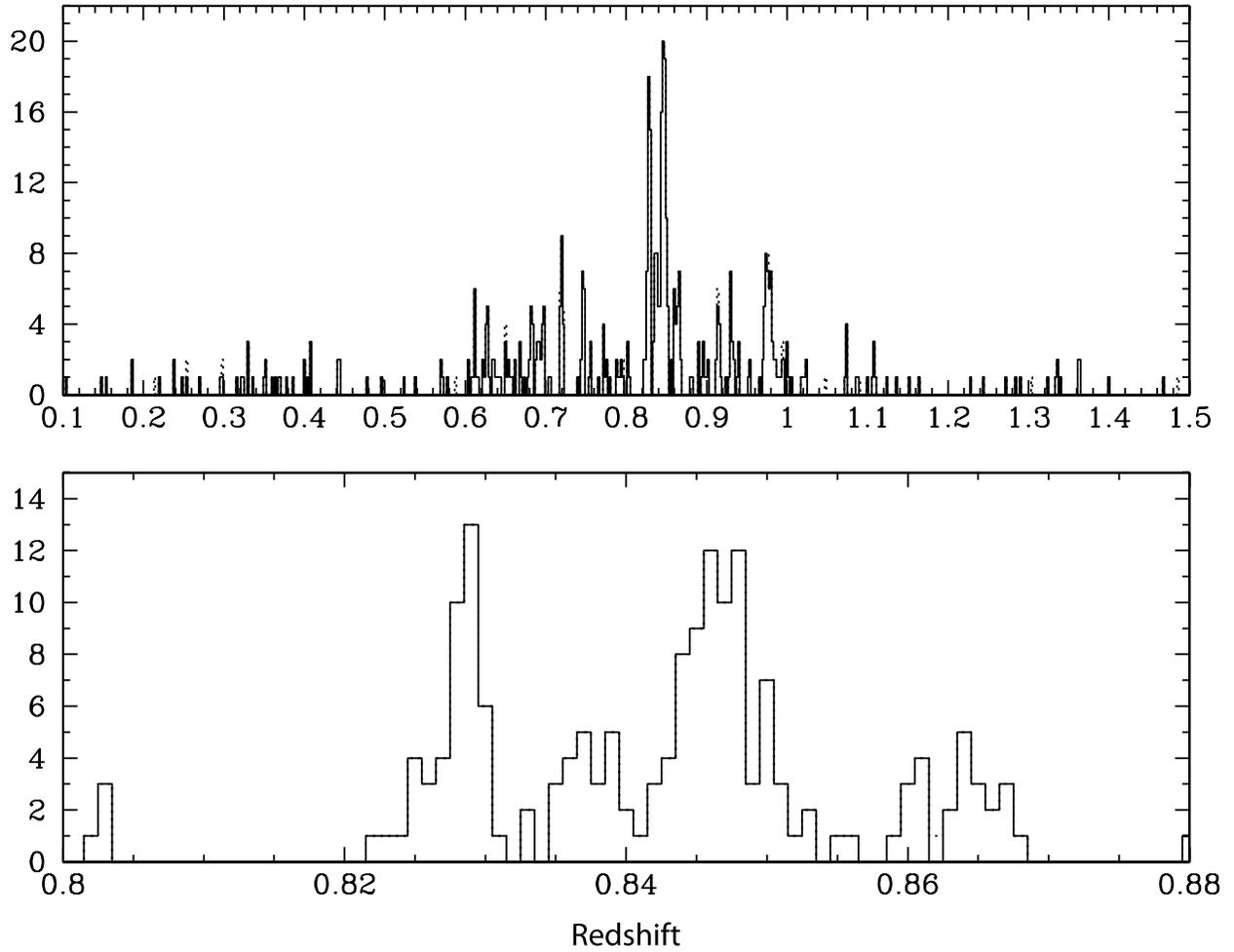}
\caption{Redshift distribution in the Cl 0023 field. The top panel
shows all extragalactic objects (dotted line) and only those with high
redshift quality (solid line). The bottom panel focuses on the
redshift range of the group-group merger with redshift bins of $\Delta
z=0.001$. There are four clear peaks in the redshift histogram,
indicative of the dynamical state of the system.}
\label{00hist} 
\end{figure}

\begin{figure}
\plotone{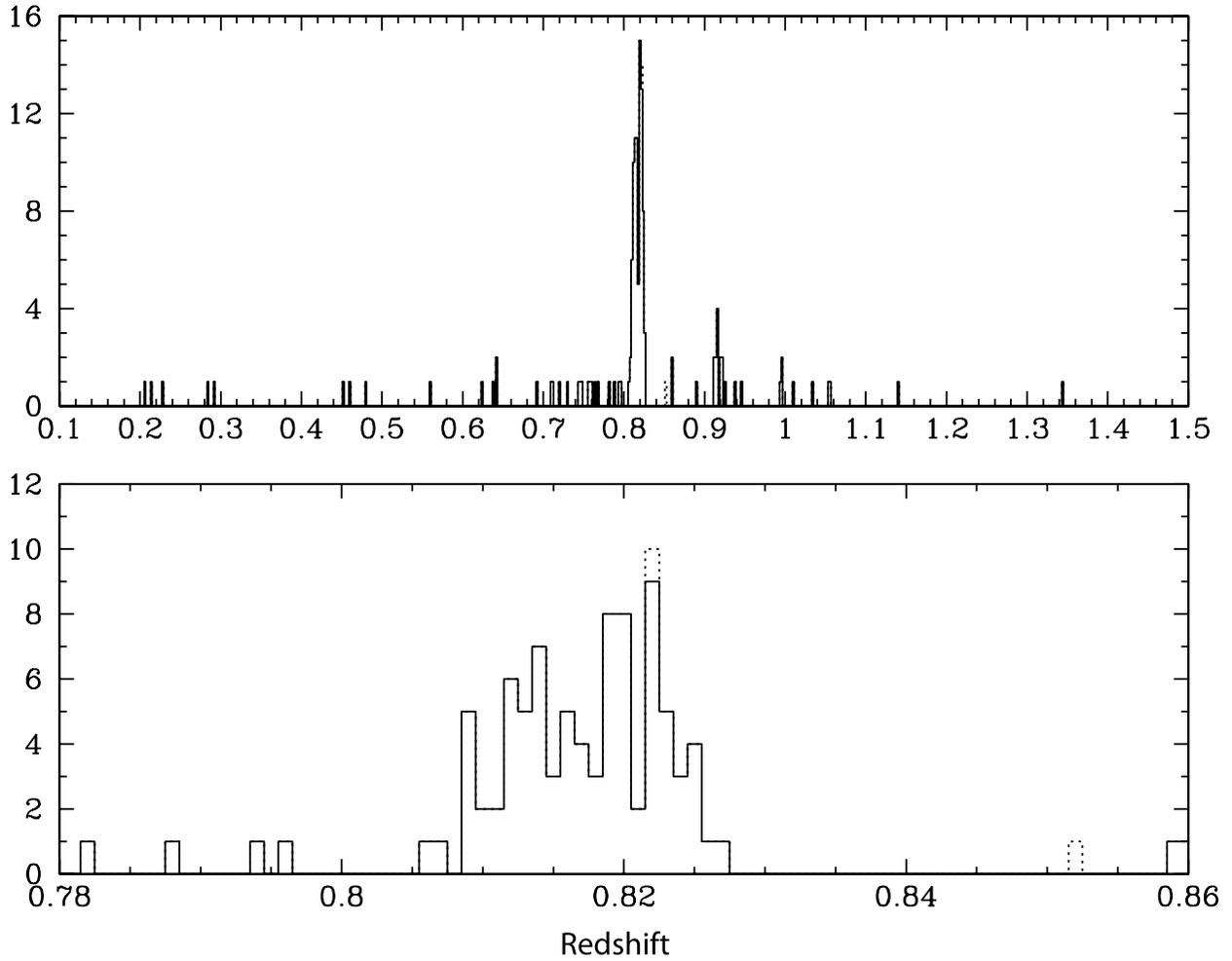}
\caption{Same as Figure~\ref{00hist} for the RX J1821 field. There is
a single clear peak $z \approx 0.817$ which is completely isolated in
redshift space; however, the distribution shows clear indications of
velocity substructure (see also Figure~\ref{52vel}).}
\label{52hist} 
\end{figure}

\begin{figure}
\plotone{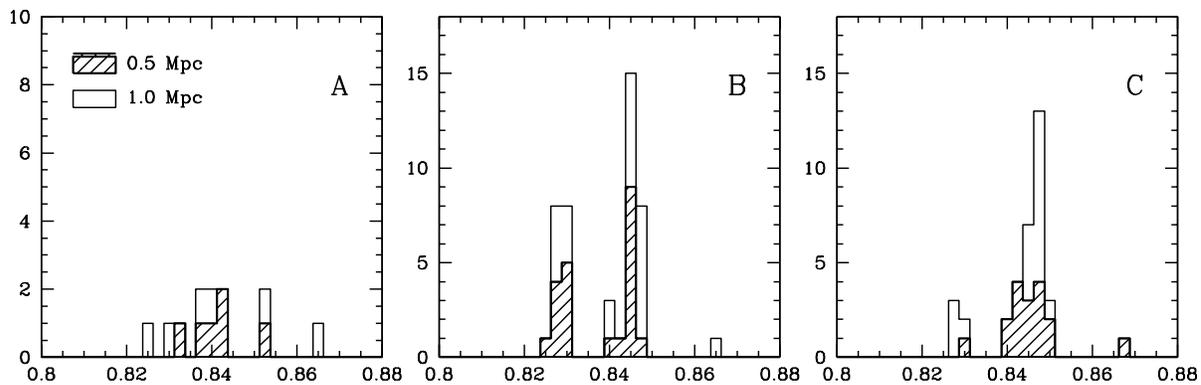}
\caption{Redshift histograms for components A, B, and C identified in
the density map of the Cl 0023 field (see Figure~\ref{00dens}). The
shaded regions show the distribution within a projected radius of 0.5
$h^{-1}_{70}$ Mpc, while the solid line corresponds to a radius of 1
$h^{-1}_{70}$ Mpc. Components A and C have reasonably well-defined
redshift peaks with measured velocity dispersions that are consistent
with galaxy groups (see Table~\ref{candinfo}), while component B is
actually a superposition of two distinct groups along the line of
sight.}
\label{00vel} 
\end{figure}

\begin{figure}
\plotone{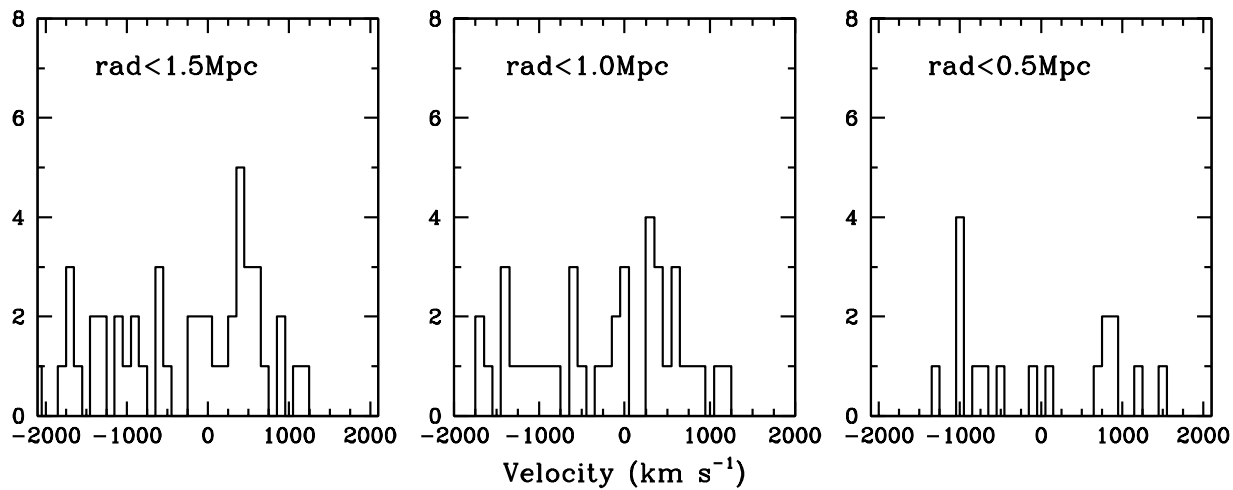}
\caption{Velocity histograms of the cluster RX J1821 within projected
radii of 0.5, 1.0, and 1.5 $h^{-1}_{70}$ Mpc. Each velocity is plotted
relative to the median velocity of the member galaxies in each
bin. Although the velocity dispersion estimators are quite consistent
within a radial bin and from bin to bin, the distributions are clearly
non-gaussian. The cluster does not, however, have clear multiple
components like Cl 0023 or a significant amount of nearby structure
like the Cl 1604 supercluster.}
\label{52vel} 
\end{figure}

\begin{figure}
\plotone{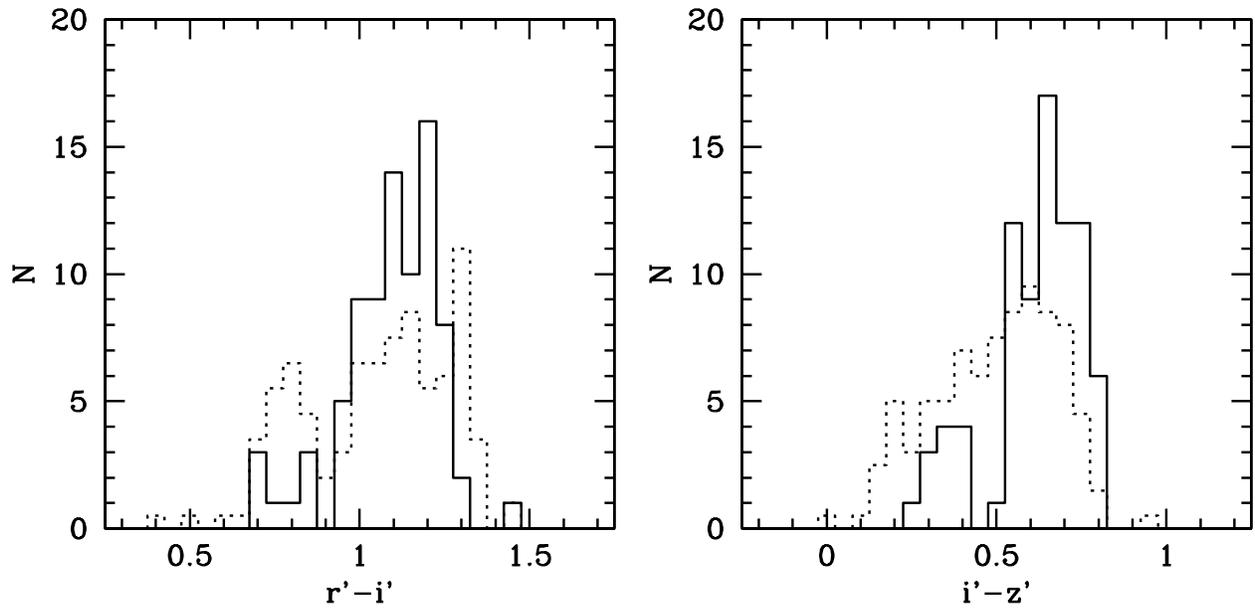}
\caption{Histogram of $r'-i'$ (left panel) and $i'-z'$ (right panel)
colors of confirmed member galaxies (down to $i'=24$) in Cl 0023
(dotted histogram) and RX J1821 (solid histogram). RX J1821 shows a
tighter red-sequence and a smaller percentage of blue galaxies than Cl
0023.}
\label{chist} 
\end{figure}

Although Cl 0023 and RX J1821 are at almost identical redshifts, they
are distinctly different in their evolutionary state. As a four-way
group merger, Cl 0023 is at the very early stage of hierarchical
cluster formation. In comparison, the cluster RX J1821 has progressed
further along its dynamical evolutionary track
as evidenced by its more concentrated velocity and spatial
structure. These global differences translate to substantial
differences in their galaxy populations as well. Based on the CMDs in
Figures~\ref{00cmd}--\ref{52cmd} and the color histograms in
Figure~\ref{chist}, Cl 0023 has a significantly larger population of
blue galaxies, with 51\% being bluer than the red-galaxy color-color
cut (down to $i' = 24.5$), compared to only 24\% in RX J1821. RX J1821
also has a tight, well-defined red-sequence which extends down to at
least $i' \simeq 23.5$, unlike the wider one of Cl 0023 (although its
larger width may be due, in part, to photometric errors; see \S
\ref{pobs}). The CMD of RX J1821 indicates a population of both old,
massive (bright) red galaxies with high formation epochs ($z_f \sim
2-3$), consistent with previous studies of massive clusters (e.g., Mei
et al.\ 2006b and references therein), as well as quenched galaxies as
evidenced by the presence of faint red-sequence galaxies (e.g., Koyama
et al.\ 2007 and references herein) and stronger H$\delta$ absorption
(see below).
 
The diversity in activity is also evident in the galaxy spectra. In Cl
0023, 80\% of the confirmed members have detectable [OII] emission
(equivalent widths greater than 2 \AA), including detectable line
emission in otherwise red, passive galaxies (with Ca H\&K but no
Balmer absorption). We note that some of the [OII] emission, in
particular in these passive galaxies, may actually be due to AGN
and/or extended emission-line regions, i.e., Seyferts or LINERs (e.g.,
Yan et al.\ 2006). This high [OII] fraction is consistent with the
results of Poggianti et al.\ (2006) who studied a sample of groups and
clusters at $z = 0.4-0.8$ from the ESO Distant Cluster Survey
(EDisCS). They find an anti-correlation between [OII] fraction
($f_{\rm [OII]}$) and velocity dispersion ($\sigma$), with the
majority of groups ($\sigma < 500$ km s$^{-1}$) having $f_{\rm [OII]}
\gtrsim 0.7$ (see Figure 4 of Poggianti et al.\ 2006). Thus, the high
[OII] fraction in Cl 0023 could just result from the fact that the
constituents of the system are all group-sized. In contrast, the
population of the massive cluster RX J1821 is more quiescent, with
only 36\% of the galaxies showing [OII] emission. The [OII] fraction
is completely consistent with that found in EDisCS clusters with
similar velocity dispersions. Although comparable to other
moderate-to-high redshift clusters, we note that the measured fraction
in RX J1821 is higher than the $f_{\rm [OII]} \approx 0.23$ found in
nearby ($z = 0.04-0.08$) SDSS clusters with similar dispersions (see
Figure 6 of Poggianti et al.\ 2006), consistent with the increase in
blue, star-forming cluster galaxies with redshift (e.g., Butcher \&
Oemler 1984; Ellingson et al.\ 2001).

\begin{figure}
\plotone{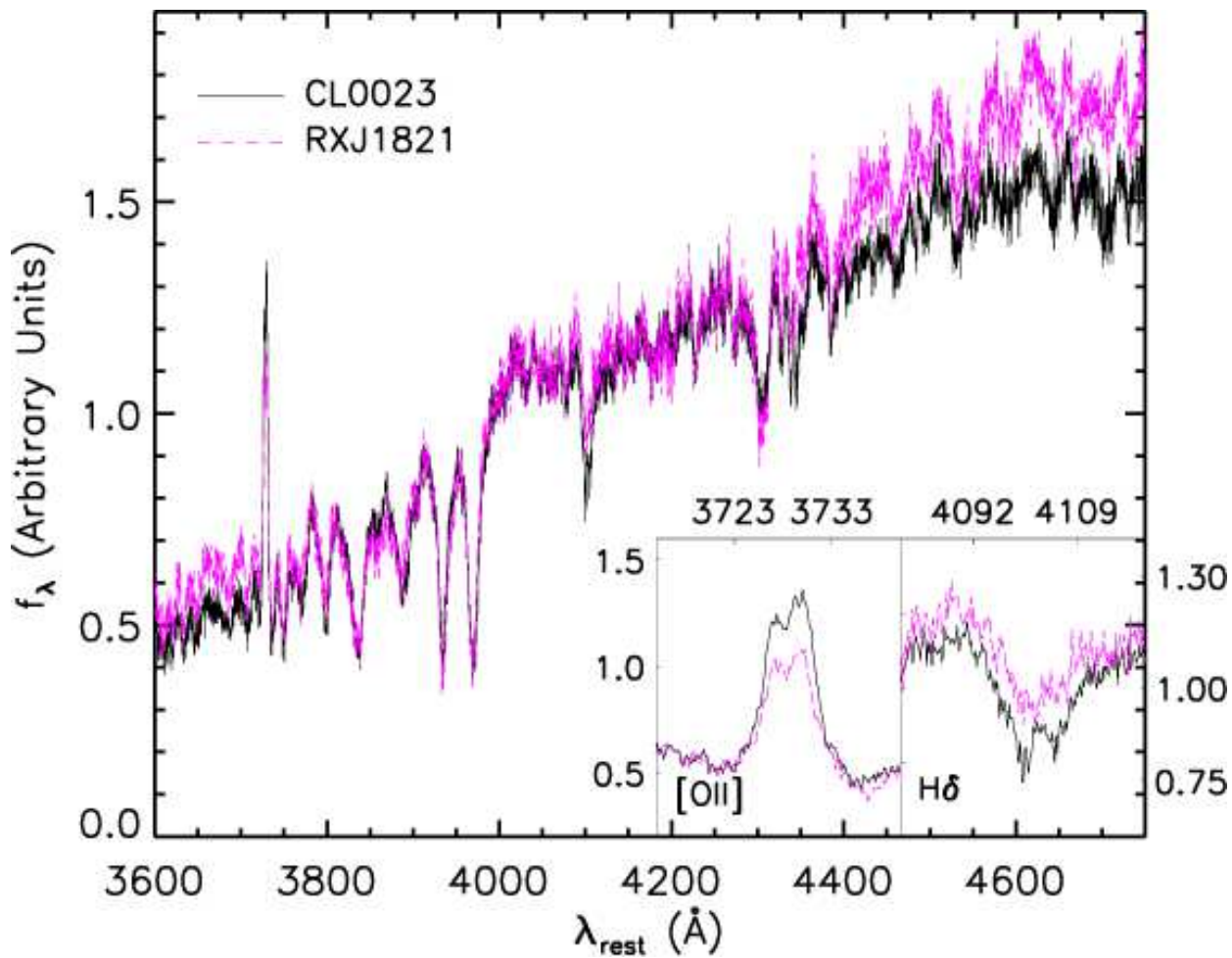}
\caption{A comparison between the composite DEIMOS spectra of Cl 0023
(solid black line) and RX J1821 (dashed magenta line). Only confirmed
members with redshift quality $Q \ge 3$ are used in both cases: 138
and 69 members for Cl 0023 and RX J1821, respectively (see \S
\ref{galaxy}). The two composites are normalized so that their fluxes
match at 4050 \AA. Insets in the lower right show close ups of the
[OII] and H$\delta$ lines. As already indicated by their colors, the
galaxy population of Cl 0023 is considerably more active than that of
RX J1821.}
\label{spec} 
\end{figure}

In Figure~\ref{spec}, we show a comparison between the composite
spectra from the Cl 0023 and RX J1821 galaxy populations. When making
the composites, we use only the DEIMOS spectra because of their higher
quality and spectral resolution. Each spectrum is normalized (by their
average rest-frame 3700--4300 \AA\ flux) and weighted by its $i'$-band
luminosity. A small fraction (less than 13\%) of galaxies are excluded
because of instrumental effects or bad columns in their spectra or
non-detections in the photometry. The resulting composites include 138
and 69 galaxies in Cl 0023 and RX J1821, respectively. The comparison
reveals clear differences between the two galaxy populations. The Cl
0023 composite has stronger [OII] and H$\delta$ lines, as expected
from the overall blue colors of the galaxy population. There is also
evidence of significant infill of the H$\delta$ line from ongoing star
formation which is essentially absent in the RX J1821 data. Without
this infill (or equivalently $\sim 100$ Myr after star formation is
truncated), the apparent contribution from A stars (i.e., the
post-starburst mode) would be much higher than for RX
J1821. Conversely, the composite of RX J1821 has strong Ca H\&K lines
and a significant excess of flux at wavelengths greater than 4500 \AA,
showing a stronger contribution from K and M stars. Using the standard
bandpasses from Fisher et al.\ (1998), we measure the equivalent
widths (EWs) of the [OII] and H$\delta$ line for each composite
spectrum. We find $\{{\rm OII, H}\delta\}$ = $\{-8.0, 3.9\}$ for Cl
0023 and $\{-4.7, 1.4\}$ for RX J1821. Using slightly different
bandpasses (from e.g., Postman et al.\ 1998 and Balogh et al.\ 1999)
or the line-fitting functionality of the {\sc iraf}\footnote{IRAF
(Image Reduction and Analysis Facility) is distributed by the National
Optical Astronomy Observatories, which are operated by the Association
of Universities for Research in Astronomy under cooperative agreement
with the National Science Foundation.} task {\it splot} changes the
EWs by less than 20\% for H$\delta$ and less than 10\% for [OII].

Because the spectroscopy in Cl 0023 is more extensive than that in RX
J1821 (i.e., 5 versus 2 slitmasks, respectively), it is possible that
we probe further into the bluer, low spectroscopic priority
population, thus skewing the comparison of the two systems. To test
this possibility, we compared only the data from the first two
slitmasks for each of the two fields. In these subsamples, Cl 0023 has
85 confirmed members, 54 of which (64\%) meet our red galaxy color
cut, whereas RX J1821 has 69 confirmed members, with 56 (81\%) meeting
this color cut. These percentages, along with
Figures~\ref{00cmd}--\ref{52cmd}, confirm the higher red galaxy
fraction in RX J1821. The excess red galaxies in RX J1821 are faint,
red sequence galaxies which are much less common in Cl 0023, where we
instead find more luminous blue galaxies. The EWs of the [OII] and
H$\delta$ lines in the Cl 0023 composite are essentially unchanged
from the full sample. Even if we consider only the high-priority (red)
galaxy populations from the first two masks, we find $\{{\rm OII,
H}\delta\}$ = $\{-5.8, 3.4\}$ for Cl 0023 and $\{-2.9, 1.0\}$ for RX
J1821. As expected, the overall line strengths are weaker in the red
galaxies than in the full sample, but the differences between the two
clusters persist. Even the red galaxy population in Cl 0023 is more
active in star-forming/AGN activity than it is in RX J1821.

\begin{figure}
\epsscale{0.8}
\plotone{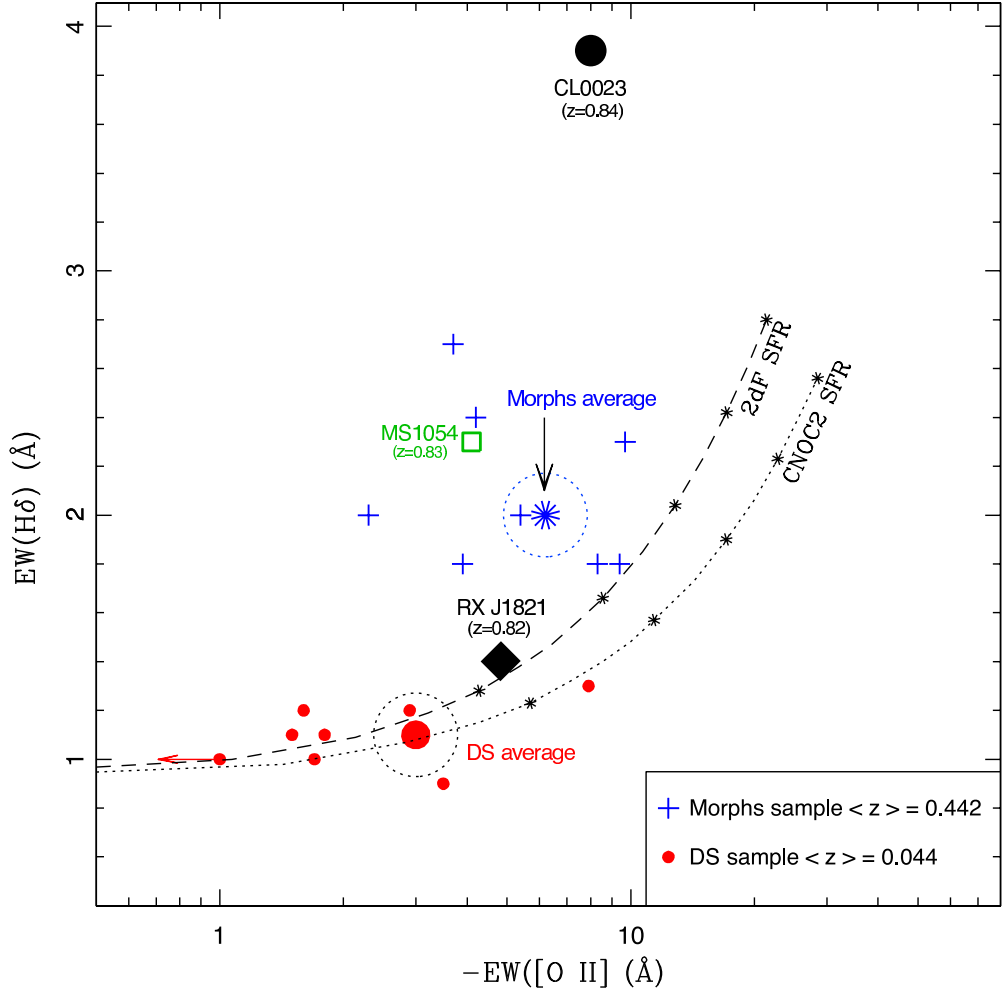}
\caption{Adapted Dressler et al.\ (2004) plot of [OII] vs.\ H$\delta$
equivalent width. The two dashed lines (based on the CNOC2 and 2dF
surveys) indicate the average spectral properties of a galaxy
population consisting of passive galaxies plus a varying fraction of
{\it continuously} star-forming galaxies. The small asterisks along
the curves mark the appropriate values of EW([OII]) and EW(H$\delta$)
for 20\%, 40\%, 60\%, 80\%, and 100\% mix of galaxies undergoing
continuous star formation. Local clusters fall on the lines, while
moderate-redshift clusters, as does the massive cluster MS 1054 at $z
= 0.83$ (open green square), lie above indicating some contribution
from galaxies that have undergone a recent starburst. The average
properties of the Cl 0023 system (large black circle) is substantially
offset in both [OII] and H$\delta$, implying a high fraction of
star-forming galaxies and an even larger contribution from recent
starbursts. In contrast, the X-ray--selected, isolated cluster RX
J1821 at $z = 0.82$ (large black diamond) has more moderate [OII]
emission and falls close to the normal star-formation lines,
indicating a large quiescent galaxy population (as also evident in the
CMD of Figure~\ref{52cmd}) and only a modest starburst contribution.}
\label{dressler} 
\end{figure}

In Figure~\ref{dressler}, we place the results of our composite
spectra (from the total galaxy samples) on the Dressler et al.\ (2004)
plot of [OII] vs.\ H$\delta$ equivalent width. The two dashed lines,
based on the CNOC2 and 2dF field surveys, indicate the average
spectral properties of a galaxy population which consists of passive
galaxies plus a varying fraction of {\it continuously} star-forming
galaxies. The low-redshift 2dF data represent the low star formation
rate (SFR) case, while the moderate-redshift ($z \sim 0.4$) CNOC2 data
represents the high SFR case. The small asterisks on the curved lines
indicate subsequently higher fractions of normal star-forming
galaxies, ranging from 20 to 100\% of the total cluster
population. Local clusters from the Dressler \& Schectman (1988;
hereafter DS) sample fall directly on the lines, indicating a cluster
population largely dominated by quiescent galaxies plus a small
contribution of continuously star-forming galaxies. Moderate-redshift
($\langle z \rangle = 0.442$) clusters from the Morphs sample
(Dressler et al.\ 1999), as well as the massive cluster MS 1054 at $z
= 0.83$ (van Dokkum et al.\ 2000), all have average H$\delta$
strengths of $\sim 2$ \AA\ and have substantial [OII] emission ranging
from --3 to --10 \AA. Both features indicate significantly more
on-going star formation than found in the local sample (see Dressler
et al.\ 2004). Unlike the nearby clusters, the moderate-redshift
clusters lie above the lines, implying an H$\delta$ line which is too
strong to be produced by normal star formation.  These results
indicate a larger contribution from galaxies that have undergone a
recent starburst.

When placed on this diagram, we see that the average properties of the
Cl 0023 population is on the high end of the [OII] distribution from
the Morphs sample and, more significantly, substantially offset in
H$\delta$, above both the normal star-forming lines as well as the
Morphs average. These results imply an even larger contribution from
recent starbursts in the Cl 0023 system. In contrast, the RX J1821
population has a weaker [OII] line, nearer to the low-end of the
Morphs sample, and falls considerably closer to the normal
star-forming lines. These results indicate a large quiescent galaxy
population, a small population of continuously star-forming galaxies
(as also evident in the CMD of Figure~\ref{52cmd}), and only a modest
contribution from starburst galaxies. 
As noted by Dressler et al.\ (2004), there is strong evolution in the
[OII] and H$\delta$ features from low-to-moderate redshift. However,
there is no obvious trend to higher redshifts. At $z \approx 0.4-0.9$,
the structures all occupy regions of phase space which corresponds to
activity levels higher than local clusters. In all cases, they lie
above the normal star-formation lines but with large
cluster-to-cluster variations, indicating different evolutionary
histories and, subsequently, different starburst contributions.

As a caveat, we note that an exact comparison of Cl 0023 and RX J1821
to other surveys depends on the spectroscopic selection criteria. We
have shown above that we can directly compare the populations of Cl
0023 and RX J1821. However, the DS and Morphs spectroscopic samples
are derived largely (although not entirely; see Dressler et al.\ 1999)
from magnitude-limited surveys. Because red (presumably less active)
galaxies are given a higher priority in our spectroscopic survey, the
resulting composite spectra of Cl 0023 and RX J1821 may be biased,
relative to the DS and Morphs samples, toward smaller [OII] and
H$\delta$ EWs. As such, our measurements would represent lower limits
on the star formation activity. In the case of Cl 0023, correcting
this bias would likely increase its already substantial offset from
the lower-redshift samples and the normal star-formation lines,
further confirming the extreme level of activity in this system.

To quantify this bias, we plan to measure the average (and range) of
spectral properties (e.g., EWs) as a function of color and magnitude
once the spectroscopic data are completed for a significant portion of
our survey. Incompleteness, especially in the faint blue population,
can be measured by statistically subtracting the typical field
contribution in any given color and magnitude range using deep imaging
taken with the same filters as part of the CFHTLS. Combining the two,
we can estimate (and account for) the effect of bias on the cluster
properties, in particular the blue fractions and composite spectra. In
addition, we will use photometric redshifts where possible to more
accurately determine our spectroscopic selection function. However,
accurate photometric redshifts require mid-infrared data in the IRAC
bands which are currently only available for the Cl 1604 supercluster
field.

\label{galaxy}

\section{Conclusions}

In this paper, we have presented the motivation, design, and
implementation of the Observations of Redshift Evolution in Large Scale
Environments (ORELSE) survey, a systematic search for structure on
scales of 10 $h^{-1}_{70}$ Mpc around 20 well-known clusters at $0.6 <
z < 1.3$. The survey covers 5 square degrees, all targeted at
intermediate-to-high-density regions, making it complementary and
comparable to field surveys such as DEEP2 and COSMOS.  The program
utilizes optical/near-infrared imaging at the UKIRT 3.8-m, KPNO 4-m,
Palomar 5-m, and Subaru 8-m covering at least 25$'$ around each target
cluster. Following the successful application in the Cl 1604
supercluster at $z \approx 0.9$, we use an adaptively smoothed
red-galaxy density map to visually identify associated groups/clusters
and larger scale filaments/walls. Guided largely by actual
observations, we use the extensive sample of 400+ confirmed members in
Cl 1604 supercluster to adapt our color and magnitude cuts, quantify
significant density peaks, and estimate false contamination rates for
other target fields at lower and higher redshifts.  We find that this
technique is highly efficient at detecting systems even down to group
masses ($\sigma \sim 300$ km s$^{-1}$), as well as extended structures
covering significant portions of the imaging field.

The crucial component of the ORELSE survey, and what distinguishes it
from similar surveys, is the unprecedented amount of spectroscopic
coverage. Utilizing the wide field, multi-object spectrograph DEIMOS
on the Keck 10-m, we are obtaining high quality spectra for 100--200+
confirmed members per system, allowing us to measure properties on a
galaxy-to-galaxy basis. Targeting galaxies down to $i' = 24$, our
system of color selection provides a spectroscopic efficiency for
cluster members of up to 45\%.

Based on the first results from the survey, we already see significant
diversity of large-scale and galaxy-scale properties at a given
redshift. Although at similar redshifts, Cl 0023 and RX J1821 differ
in their evolutionary state, global dynamics, activity level, and
galaxy population. The optically-selected Cl 0023 shows multiple
high-density peaks in the red-galaxy density map and a complex
velocity structure, suggesting a four-way group-group merger. The
measured velocity dispersions of the member groups range from 206--479
km s$^{-1}$. The overall galaxy population is exceptionally active,
with a high fraction of blue galaxies and [OII] emitters. Conversely,
the X-ray selected RX J1821 is a relatively isolated,
centrally-condensed massive cluster. Only one significant overdensity
and one redshift peak are observed in this field. The measured
velocity dispersion is 926 km s$^{-1}$, indicative of a rich
cluster. However, there are obvious signs of continuing dynamical
evolution, including an elongated galaxy distribution and significant
velocity substructure. The galaxy population is substantially more
quiescent than that of Cl 0023; however, it still has a larger
fraction of blue, star-forming galaxies than local clusters,
consistent with the bluing of the cluster population with
redshift. Composite spectra from both systems suggest some
contribution from galaxies which have undergone a recent starburst
(strong H$\delta$ absorption), with the contribution in Cl 0023 being
substantial.

These two systems indicate that global cluster properties (i.e., mass)
play a role in driving galaxy evolution (see also Discussion of
Poggianti et al.\ 2006). Cl 0023 is comprised of group-sized halos,
has a small quiescent population, and subsequently contains more
active galaxies. Cl 0023 will likely be a massive cluster, with a
velocity dispersion on the order of RX J1821, in $\sim 1$ Gyr,
although noticeable dynamical evolution will still be taking place
after $\sim 3$ Gyr, i.e.\ until $z \lesssim 0.3$ (Lubin, Postman \&
Oke 1998). If this dynamical evolution does take place, the star
formation in a significant fraction of the galaxy population must be
quenched during this period in order to be consistent with average
star-formation rates and blue fractions in moderate-redshift clusters
of similar mass (e.g., Balogh et al.\ 1997; Nakata et al.\ 2005). The
current activity level in Cl 0023 and this evolutionary time frame
implies that the group and/or merger environment is efficient at
inducing and quenching star formation.  Conversely, RX J1821 is
already a cluster-sized halo, with a galaxy population dominated by
passive red galaxies (from both early-epoch formation and later-time
quenching). Although both systems will end up as massive clusters by
the present day, they have clearly different evolutionary
histories. We note that these results, along with preliminary findings
from other ORELSE targets, suggest that X-ray--selection may be
biased, at least with respect to optical surveys, toward more evolved
systems at a given redshift. On the other hand, due to projection
effects, optical surveys detect multi-group systems and systems
embedded in large scale structures that are not traditionally part of
``cluster'' surveys. As a result, the cluster detection technique may
have a strong impact on any conclusions concerning the presence and
effect of large scale structure, the relative importance of global
versus local environment, and the timescales for galaxy and cluster
evolution.

The scientific goals of the ORELSE survey are to identify and examine
a statistical sample of dynamically active clusters and large scale
structures during an active period in their history. Based on our
completed (including spectroscopic) observations so far, such as those
presented here and in Gal et al.\ (2008), we have found two
superclusters (containing 7+ groups/clusters), a four-way group-group
merger, and two largely-isolated, massive X-ray--selected
clusters. When the survey is complete, we will have a significant
sample with which to constrain large scale cluster dynamics, cluster
formation mechanisms, and the effect of environment on galaxy
evolution.  Combining our ground-based databases with follow-up radio,
X-ray, and/or mid-infrared observations, the program will (1)
determine the relation between global cluster properties (such as
richness, mass, luminosity, and dynamical state) and the nature of the
surrounding large scale structures; (2) quantify the dependence of
galaxy multi-wavelength and spectral properties on position and
density to determine when, where, and why star formation and nuclear
activity are triggered (or truncated) in and near the cluster
environment; and (3) chart the process of structure formation and its
effect on galaxy evolution over this active 3 Gyr period.

\acknowledgements

We would like to thank the anonymous referee for very constructive
comments on this manuscript. This material is based upon work
supported by the National Aeronautics and Space Administration under
Award No.\ NNG05GC34ZG for the Long Term Space Astrophysics
Program. The spectrographic data presented herein were obtained at the
W.M. Keck Observatory, which is operated as a scientific partnership
among the California Institute of Technology, the University of
California and the National Aeronautics and Space Administration. The
Observatory was made possible by the generous financial support of the
W.M. Keck Foundation. The authors wish to recognize and acknowledge
the very significant cultural role and reverence that the summit of
Mauna Kea has always had within the indigenous Hawaiian community.  We
are most fortunate to have the opportunity to conduct observations
from this mountain. The observing staff, telescope operators, and
instrument scientists at Keck provided a great deal of assistance.

\clearpage

\begin{deluxetable}{lcclcccclc}
\rotate
\setlength{\tabcolsep}{0.05in}
\tablewidth{585pt}
\tablecaption{The Cluster Sample}
\tabletypesize{\scriptsize}      
\tablehead{
\colhead{Cluster}&  \colhead{R.A.\ (J2000)} & \colhead{Dec.\ (J2000)} & \colhead{$\bar{z}$} & \colhead{Selection} & \colhead{Chandra/XMM} & \colhead{Optical} & \colhead{Near-infrared} & Notes & Refs\\
\colhead{} & \colhead{} & \colhead{} & \colhead{} & \colhead{} &  \colhead{Observations?} & \colhead{Complete?} & Complete?}
\startdata
Cl 0023+0423    & 00 23 52.2    & +04 23 07   & 0.845        & Optical & Yes & Yes & Partial & complex merger & (1) \\
RCS J0224-0002  & 02 24 30.0    & $-$00 02 00 & 0.772        & Optical & Yes &  &  & strong lens & (2) \\
XLSSC005a+b     & 02 27 09.7    & $-04$ 18 05 & 1.000        & X-ray   & Yes & Yes\tablenotemark{a} &  & double cluster & (3) \\
Cl J0849+4452   & 08 48 56.3    & +44 52 16   & 1.261        & X-ray   & Yes & Yes\tablenotemark{b} &  & Lynx double cluster & (4) \\
Cl J0910+5422   & 09 10 44.9    & +54 22 09   & 1.110        & X-ray   & Yes & Yes & Yes & elongated structure & (5) \\
Cl 0934+4804    & 09 43 41.0    & +48 04 46   & 0.699        & Optical & Yes & Yes & Yes & & (6) \\
RX J1053.7+5735 & 10 53 39.8    & +57 35 18   & 1.140        & X-ray   & Yes & Yes\tablenotemark{b}  & Yes & double cluster & (7) \\
1137+3000       & 11 37 33.4    & +30 07 36   & 0.959        & Radio   &     & Yes\tablenotemark{b}  & Yes & wide-angle-tailed radio source & (8) \\
RX J1221.4+4918 & 12 21 24.5    & +49 18 13   & 0.700        & X-ray   & Yes & Yes & Yes & double cluster & (9) \\
Cl 1324+3011    & 13 24 50.4    & +30 11 26   & 0.756        & Optical & Yes & Yes & Yes & supercluster member & (10) \\
Cl 1324+3059    & 13 24 53.5    & +30 59 00   & 0.755        & Optical & Yes & Yes & Yes & supercluster member & (11) \\
Cl 1325+3009    & 13 25 18.7    & +30 09 57   & 1.067        & Optical &     & Yes\tablenotemark{b} & Yes & & (12) \\
Cl J1350+6007   & 13 50 48.5    & +60 07 07   & 0.804        & X-ray   & Yes & Yes & Yes & disturbed, elliptical structure & (13) \\
Cl J1429.0+4241 & 14 29 06.4    & +42 41 10   & 0.920        & X-ray   & Yes & Yes & Yes & elongated structure & (14) \\
Cl 1604+4304    & 16 04 19.5    & +43 04 33   & 0.900        & Optical & Yes & Yes & Yes & supercluster member & (15) \\
Cl 1604+4321    & 16 04 31.5    & +43 21 17   & 0.920        & Optical & Yes & Yes & Yes & supercluster member & (16) \\
RX J1716.4+6708 & 17 16 49.6    & +67 08 30   & 0.813        & X-ray   & Yes & Yes\tablenotemark{b} & Yes & elongated core, small subcluster & (17) \\
RX J1757.3+6631 & 17 57 19.4    & +66 31 31   & 0.691        & X-ray   & Yes & Yes & Yes & & (18) \\
RX J1745.2+6556 & 17 45 18.2    & +65 55 42   & 0.608        & X-ray   & & Yes & Yes & & (19) \\
RX J1821.6+6827 & 18 21 32.9    & +68 27 55   & 0.818        & X-ray   & Yes & Yes & Yes & elongated structure &(20) \\
\enddata
\tablenotetext{a}{Arhival data from the CFHT Legacy Survey (see {\tt http://www.cfht.hawaii.edu/Science/CFHTLS/}).}
\tablenotetext{b}{Arhival data from Subaru (see {\tt http://smoka.nao.ac.jp/}).}
\tablerefs{(1) Gunn, Hoessel \& Oke (1986); Oke, Postman, \& Lubin (1998); 
(2) Gladders \& Yee (2000); Gladders, Yee, \& Ellingson (2002); 
(3) Valtchanov et al.\ (2004); Andreon et al.\ (2005); Pierre et al.\ (2006);
(4) Rosati et al.\ (1998, 1999); Nakata et al.\ (2005); Mei et al.\ (2006b);
(5) Stanford et al.\ (2002); Mei et al.\ (2006a); 
(6) Gunn, Hoessel, \& Oke (1986); Oke, Postman, \& Lubin (1998);
(7) Hashimoto et al.\ (2005);
(8) Blanton et al.\ (2003); 
(9) Vikhlinin et al.\ (1998); Mullis et al.\ (2003); Jeltema et al.\ (2005);
(10) Gunn, Hoessel, \& Oke (1986); Oke, Postman, \& Lubin (1998);
(11) Gunn, Hoessel, \& Oke (1986); Oke, Postman, \& Lubin (1998);  
(12) Postman et al.\ (1996); Oke, Postman, \& Lubin (1998);
(13) Rosati et al.\ (1998); Holden et al.\ (2002); Jeltema et al.\ (2005);
(14) Maughan et al.\ (2006);
(15) Gunn, Hoessel, \& Oke (1986); Oke, Postman, \& Lubin (1998);
(16) Gunn, Hoessel, \& Oke (1986); Oke, Postman, \& Lubin (1998);
(17) Gioia et al.\ (1999); Jeltema et al.\ (2005);
(18) Gioia et al.\ (2003); Henry et al.\ (2006);
(19) Gioia et al.\ (2003); Henry et al.\ (2006);
(20) Gioia et al.\ (2003, 2004); Henry et al.\ (2006)}
\label{sample}
\end{deluxetable}

\clearpage

\begin{deluxetable}{lll}
\tablecaption{The Red-Galaxy Color Selection}
\tabletypesize{\scriptsize}      
\tablecolumns{16}
\tablewidth{0pc}
\tablehead{
\colhead{Redshift Range}&  \colhead{$r'-i'$} & \colhead{$i'-z'$}}
\startdata
$0.6-0.7$ & $1.0-1.4$ & $0.3-0.7$ \\
$0.7-0.8$ & $1.0-1.4$ & $0.4-0.8$ \\
$0.8-0.9$ & $1.0-1.4$ & $0.5-0.9$ \\
$0.9-1.0$ & $1.0-1.4$ & $0.6-1.0$ \\
$> 1.0$ & \nodata & $> 0.9$  \\
\enddata
\label{ctab}
\end{deluxetable}

\clearpage

\begin{deluxetable}{llrrcrrlrrrl}
\tabletypesize{\scriptsize}
\tablecolumns{16}
\tablewidth{0pc}
\tablecaption{Target Coordinates, Redshifts, and Velocity Dispersions}
\tablehead{
\colhead{}  & \colhead{}  &  \multicolumn{2}{c}{J2000.0} & \colhead{} & \multicolumn{3}{c}{Within $0.5 h^{-1}_{70}$ Mpc} & \colhead{} &  \multicolumn{3}{c} {Within $1.0 h^{-1}_{70}$ Mpc}\\
\cline{3-4} \cline{6-8} \cline{10-12} \\[-6pt]
\colhead{Name} & \colhead{ID} & \colhead{RA} & \colhead{Dec} & \colhead{} & \colhead{$N$} & \colhead{$z_{mean}$}  
& \colhead{$\sigma$} & \colhead{} & \colhead{$N$} & \colhead{$z_{med}$} & \colhead{$\sigma$}}
\startdata
Cl 0023+0423 & A & 6.0223 & 4.3586 & & 5 & 0.8383 & \tablenotemark{a}\nodata & & 7 & 0.8390 & 479 $\pm$ 172 \\
& B1\tablenotemark{b} & 5.9764 & 4.3874 & & 10 & 0.8286 & 242 $\pm$ 54  & & 17 & 0.8282 &  206 $\pm$ 32 \\
& B2\tablenotemark{b} & 5.9671 & 4.3824 & & 12 & 0.8445 & 177 $\pm$ 118 & & 27 & 0.8453 &  293 $\pm$ 79 \\
& C  & 5.9227 & 4.3824 & & 15 & 0.8451 & 548 $\pm$ 84  & & 29 & 0.8463 &  428 $\pm$ 67 \\
RX J1821.6+6827 & & 275.3801 & 68.4651 & & 17 & 0.8170 & 993 $\pm$ 87 & & 40 & 0.8193 & 926 $\pm$ 77 \\
\enddata
\tablenotetext{a}{Insufficient redshifts to compute dispersion} \tablenotetext{b}{There are two distinct velocity
peaks in this component (see Figure~\ref{00vel}).}
\label{candinfo}
\end{deluxetable}

\clearpage

\clearpage

\clearpage

\clearpage

\clearpage

\clearpage

\clearpage

\clearpage

\clearpage

\clearpage

\clearpage

\end{document}